\documentclass[runningheads]{llncs}
\usepackage{llncsdoc}
\usepackage{epsfig}
\usepackage{color}
\begin{document}
\title{
Symbolic-numerical Algorithm for Generating\\ Cluster
Eigenfunctions: Tunneling of Clusters Through Repulsive Barriers
         }
\titlerunning{Tunneling of Clusters Through Repulsive Barriers}
\author{
Sergue Vinitsky\inst{1},
Alexander Gusev\inst{1},
Ochbadrakh Chuluunbaatar\inst{1},\\
Vitaly Rostovtsev\inst{1},
Luong Le Hai\inst{1,2},
Vladimir Derbov\inst{3},
Pavel Krassovitskiy\inst{4}
}

\authorrunning{Sergue Vinitsky\ et al.}
\institute{
Joint Institute for Nuclear Research, Dubna, Moscow Region, Russia e-mail:
\email{vinitsky@theor.jinr.ru}
\and
Belgorod State University, Belgorod, Russia
\and
Saratov State University, Saratov, Russia
\and
Institute of Nuclear Physics, Almaty, Kazakhstan}

\tocauthor{}
\maketitle
\begin{abstract}
 A model for quantum tunnelling of
 a cluster comprising
 $A$ identical particles,
 coupled by oscillator-type potential, through short-range repulsive potential barriers
 is introduced  for the first time in the
 new symmetrized-coordinate representation and studied
 within the s-wave approximation.
 The symbolic-numerical algorithms for calculating the effective potentials of the close-coupling equations in terms of the cluster wave functions and the
energy of the barrier quasistationary states are formulated and implemented using the Maple computer algebra system.
The effect of quantum transparency, manifesting itself in nonmonotonic resonance-type
dependence of the transmission coefficient upon the energy of the particles,
the number of the particles $A=2,3,4$, and their symmetry type, is analyzed.
It is shown that
the resonance behavior of the total transmission coefficient is due to the
 existence of barrier quasistationary states imbedded in the continuum.
  \footnote{The talk  presented at the 15th  International Workshop
''Computer Algebra in Scientific Computing 2013'', Berlin,
Germany, September 9-13, 2013}.
\end{abstract}

\section{
Introduction } During a decade, the mechanism of quantum
penetration of two bound particles through repulsive barriers
\cite{P00} attracts attention from both theoretical and
experimental viewpoints in relation with such problems  as
near-surface quantum diffusion of molecules
~\cite{Fasolino07,Ivanov10,Shegelski12}, fragmentation in
producing very neutron-rich light nuclei ~\cite{Ershov,Nesterov},
and heavy ion collisions through multidimensional barriers
\cite{Hofmann,Krappe,CwiokDudek,Hagino,Zagrebaev,Volya,Shotter11,Shilov}.
Within the general formulation of the scattering problem for ions
having different masses,  a benchmark model with long-range
potentials was proposed in Refs.~\cite{yaf09,casc11,stankin11}.
The generalization of the two-particle model over a quantum system
of $A$ identical particles is of great importance for the
appropriate description of molecular and heavy-ion collisions.
{\it The aim of this paper is to present the convenient
formulation of the problem stated above and the calculation
methods, algorithms, and programs for solving this problem.}

We consider a \textit{new method} for the description of the
penetration of $A$ identical quantum  particles, coupled by
short-range oscillator-like interaction, through a repulsive
potential barrier. We assume that the spin part of the wave
function is known, so that only the spatial part of the wave
function is to be considered, which may be symmetric or
antisymmetric with respect to a permutation of $A$ identical
particles. The initial problem is reduced to the penetration of a
composite system with the internal degrees of freedom, describing
an $(A-1)\times d$-dimensional oscillator, and the external
degrees of freedom describing the center-of-mass motion of $A$
particles in $d$-dimensional Euclidian space. For simplicity, we
restrict our consideration to the so-called $s$-wave approximation
\cite{P00} corresponding to one-dimensional Euclidean space
($d=1$).

We seek for the solution in the form of  Galerkin expansion in
terms of cluster functions in the \textit{new symmetrized
coordinate representation} (SCR) \cite{cascnedve1} with unknown
coefficients having the form of matrix functions of the
center-of-mass variable. As a result, the problem is reduced to a
boundary-value problem for a system of ordinary second-order
differential equations with respect to the center-of-mass
variable. Conventional asymptotic boundary conditions involving
unknown amplitudes of reflected and transmitted waves are imposed
on the desired matrix solution. Solving the problem was
implemented as a complex of the \textit{symbolic-numeric
algorithms and programs} in CAS MAPLE and  FORTRAN environment.
The results of calculations are analyzed with particular emphasis
on the effect of quantum transparency that manifests itself as
nonmonotonic energy dependence of the transmission coefficient due
to resonance tunnelling of the bound particles in S (A) states
through the repulsive potential barriers.

The paper is organized as follows. In Section 2, we present the
problem statement  in symmetrized coordinates. In Section 3, we
introduce the SCR of the cluster functions of the considered
problem and the asymptotic boundary conditions involving unknown
amplitudes of reflected and transmitted waves. In Section 4, we
formulate the boundary-value problem for the  close-coupling
equations in the Galerkin form using the SCR. In Section 5, we
analyze the results of numerical experiment on the resonance
transmission of a few coupled identical particles in S(A) states,
whose energies coincide with the resonance eigenenergies of the
barrier quasi-stationary states embedded in the continuum. In
Conclusion, we  sum up the results and discuss briefly the
perspectives of application of the developed approach.

\section{Problem Statement}
We consider a system of $A$ identical quantum particles having the
mass $m$ and a set of the Cartesian coordinates $x_i\in {\bf
R}^{d}$ in $d$-dimensional Euclidian space, considered as vector
$\tilde {\bf x}=(\tilde x_1,...,\tilde x_A)\in {\bf R}^{A\times
d}$ in   $A\times d$-dimensional configuration space. The
particles are coupled by the pair potentials $\tilde
V^{pair}(\tilde x_{ij})$  depending upon the relative coordinates,
$\tilde x_{ij}=\tilde x_i-\tilde x_j$, similar to a harmonic
oscillator potential $\tilde V^{hosc}(\tilde
x_{ij})=\frac{m\omega^2}{2}(\tilde x_{ij})^2$ with the frequency
$\omega$. The resulting clusters are subject to the influence of
the potentials $\tilde V(\tilde x_i)$ describing the  external
field of a target. The appropriate Schr\"odinger equation takes
the form
\begin{eqnarray*}\left[\!-\!\frac{\hbar^2}{2m}\sum_{i=1}^A\frac{\partial^2}{\partial \tilde x_i^2}
\!+\!\sum_{i,j=1; i<j}^A \tilde V^{pair}(\tilde x_{ij}) 
\!+\!\sum_{i=1}^A\tilde V(\tilde x_i)
\!-\!\tilde E\right]
\tilde\Psi(\tilde {\bf x})\!=\!0,
\end{eqnarray*}
where $\tilde E$ is the total energy of the system of $A$
particles, and $\tilde P^2={2m\tilde E}/{\hbar^2}$, $\tilde P$ is
the total momentum of the system, and $\hbar$ is Planck constant.
Using the oscillator units
$x_{osc}=\sqrt{\hbar/(m\omega\sqrt{A})}$,
$p_{osc}=\sqrt{(m\omega\sqrt{A})/\hbar}=x_{osc}^{-1}$, and
$E_{osc}=\hbar\omega\sqrt{A}/2$ to introduce the  dimensionless
coordinates $x_i=\tilde x_i/x_{osc}$, $x_{ij}=\tilde
x_{ij}/x_{osc}=x_i-x_j$, $E=\tilde E/E_{osc}=P^2$, $P=\tilde
P/p_{osc}=\tilde Px_{osc}$, $V^{pair}(x_{ij})=\tilde
V^{pair}(x_{ij} x_{osc})/E_{osc}$, $V^{hosc}(x_{ij})=\tilde
V^{hosc}(x_{ij} x_{osc})/E_{osc}=\frac{1}{A}(x_{ij})^2$ and
$V(x_i)=\tilde V(x_i x_{osc})/E_{osc}$, one can rewrite the above
equation in the form
\begin{eqnarray}\left[\!-\!\sum_{i=1}^A\frac{\partial^2}{\partial x_i^2}
\!+\!\!\sum_{i,j=1; i<j}^A \frac{1}{A}(x_{ij})^2
\!+\!\!\sum_{i,j=1; i<j}^A\!U^{pair}(x_{ij})
\!+\!\!\sum_{i=1}^A V(x_i)\!-\! E\right]
\Psi( {\bf x})\!=\!0,
\label{mo1}
\end{eqnarray}
where $U^{pair}(x_{ij})=V^{pair}(x_{ij})-V^{hosc}(x_{ij})$, i.e., if  $V^{pair}(x_{ij})=V^{hosc}(x_{ij})$,
then $U^{pair}(x_{ij})=0$.

The problem of tunnelling of a cluster of $A$ identical particles in the symmetrized coordinates
$(\xi_0,{\mbox{\boldmath $\xi$}})$, where ${\mbox{\boldmath $\xi$}}=\{\xi_1,...,\xi_{A-1}\}$:
\begin{eqnarray}\label{mo2}
\!\!\xi_0=\frac{1}{\sqrt{A}}\left(\sum_{t=1}^{A} x_t \right),
\,\,\xi_s=\frac{1}{\sqrt{A}}\left(x_1+\!\!\sum_{t=2}^{A} a_0x_t+\sqrt{A}x_{s+1}\right),
\,\,s=1,...,A-1,
\end{eqnarray}
in terms of total potential $U(\xi_0,{\mbox{\boldmath
$\xi$}})=V(\xi_0,{\mbox{\boldmath
$\xi$}})+U^{eff}(\xi_0,{\mbox{\boldmath $\xi$}})$ reads as
\cite{cascnedve1}
\begin{eqnarray}\left[-\frac{\partial^2}{\partial \xi_0^2}
+\sum_{i=1}^{A-1}\left(-\frac{\partial^2}{\partial \xi_i^2}
+ (\xi_i)^2\right)
+U(\xi_0,{\mbox{\boldmath $\xi$}})
- E\right] \Psi(\xi_0,{\mbox{\boldmath $\xi$}}) =0, \label{mo5}
\\ \nonumber
U^{eff}(\xi_0,{\mbox{\boldmath $\xi$}})=\sum_{i,j=1; i<j}^A  U^{pair}(x_{ij}({\mbox{\boldmath $\xi$}})),\quad
V(\xi_0,{\mbox{\boldmath $\xi$}})=\sum_{i=1}^A V(x_i(\xi_0,{\mbox{\boldmath $\xi$}})),
\end{eqnarray}
which is invariant under permutations $\xi_i\leftrightarrow \xi_j$
at $i,j=1,...,A-1$, i.e., the invariance of Eq. (\ref{mo1}) under
permutations $x_i\leftrightarrow x_j$ at $i,j=1,...,A$ survives
the transformation.

\section{Cluster Functions and Asymptotic Boundary Conditions}
For simplicity we restrict our consideration to the so-called
$s$-wave approximation \cite{P00}, i.e., one-dimensional Euclidian
space ($d=1$). Cluster functions
$\tilde\Phi_{j}(\xi_{0},{\mbox{\boldmath $\xi$}}),$  where
${\mbox{\boldmath $\xi$}}=\{\xi_1,...,\xi_{A-1}\},$ corresponding
to the threshold energies $\tilde \epsilon_j(\xi_{0})$  dependent
on $\xi_{0}$ as a parameter, are solutions of the parametric
eigenvalue problem {\footnotesize \begin{eqnarray}\label{de13f}
\left( -\frac{\partial^2}{\partial    {\mbox{\boldmath $\xi$}}^2}+
{\mbox{\boldmath $\xi$}}^2 +U^{}(\xi_0,{\mbox{\boldmath $\xi$}})
-\tilde \epsilon_j(\xi_{0})
\right)\tilde\Phi_{j}(\xi_0,{\mbox{\boldmath $\xi$}})=0,
\int_{-\infty}^{+\infty}\!\!\!\!\!\!\tilde\Phi_i(\xi_0,{\mbox{\boldmath
$\xi$}})\tilde\Phi_{j}(\xi_0,{\mbox{\boldmath
$\xi$}})d^{A-1}{\mbox{\boldmath $\xi$}} =\delta_{ij},
\end{eqnarray}}
where $U(\xi_0,{\mbox{\boldmath $\xi$}})=V(\xi_0,{\mbox{\boldmath
$\xi$}})+U^{eff}(\xi_0, {\mbox{\boldmath $\xi$}})$ is the total
potential that enters Eq. (\ref{mo5}). The effective potential
$U^{eff}(\xi_0,{\mbox{\boldmath $\xi$}})$ can be approximated also
by the deformed Wood--Saxon potential in the single-particle
oscillator approximation \cite{CwiokDudek}. We seek for the
cluster functions $\Phi_i(\xi_0,{\mbox{\boldmath $\xi$}})$ in the
form of an expansion over the eigenfunctions
$\Phi_{j'}^{S(A)}(\mbox{\boldmath $\xi$})$, symmetric (S) or
antisymmetric (A) with respect to a permutation of the initial $A$
Cartesian coordinates of $A$ identical particles. These functions
correspond to eigenenergies $E^{S(A)}_i$ of the
$(A-1)$-dimensional oscillator, generated by the algorithm SCR
\cite{cascnedve1}, with unknown coefficients
$\tilde\alpha^{(i)}_{j'}(\xi_{0})$:
\begin{eqnarray}
\tilde\Phi_i(\xi_{0},\mbox{\boldmath $\xi$})=\sum_{j'=1}^{j'_{\max}}
\tilde\alpha^{(i)}_{j'}(\xi_{0})\Phi^{S(A)}_{j'}(\mbox{\boldmath $\xi$}).
\label{nalpha}
\end{eqnarray}
Thus, the eigenvalue problem (\ref{de13f}) is reduced to a
linearized version of the Hartree--Fock algebraic eigenvalue
problem {\footnotesize
\begin{eqnarray}\sum_{j'=1}^{j'_{\max}}\left(\delta_{ij'}E_i^{S(A)}+U^{}_{ij'}(\xi_0)-\delta_{ij'}\tilde
\epsilon_i(\xi_{0})\right)\tilde\alpha^{(i)}_{j'}(\xi_{0})=0, ~~
\sum_{j'=1}^{j'_{\max}}\tilde\alpha^{(i')}_{j'}(\xi_{0})\tilde\alpha^{(i)}_{j'}(\xi_{0})=\delta_{ii'},
\label{noparam}
\end{eqnarray}}
where the potentials $U^{pair}_{ij'}$ and $V_{ij'}(\xi_0)$ are
expressed in terms of the integrals
\begin{eqnarray}\label{j}
U^{pair}_{ij'}=\int d^{A-1}\mbox{\boldmath $\xi$} \Phi_i^{S(A)}(\mbox{\boldmath $\xi$})
U^{eff}({\mbox{\boldmath $\xi$}})
\Phi_{j'}^{S(A)}(\mbox{\boldmath $\xi$}),\\
V_{ij'}(\xi_0)=\int d^{A-1}\mbox{\boldmath $\xi$}
\Phi_i^{S(A)}(\mbox{\boldmath $\xi$})\left(\sum_{k=1}^A V(x_k(\xi_0,\mbox{\boldmath $\xi$}))\right)\Phi_{j'}^{S(A)}(\mbox{\boldmath $\xi$}).\label{integrals}
\end{eqnarray}
The parametric \textit{algorithm SCR}, i.e.,   \textit{algorithm
PSCR}, for solving the above parametric eigenvalue problem
was implemented by means of subroutines \cite{prog07,Bunge01}, or in the single-particle approximation
by means of the subroutine  \cite{CwiokDudek} in CAS MAPLE and  FORTRAN environment.\\
(G) If $U_{ij'}(\xi_0)=U^{pair}_{ij'}$ are independent on $\xi_0$,
then $\tilde \epsilon_i(\xi_0)=\tilde\epsilon_i$ and
$\tilde\alpha^{(i)}_{j'}(\xi_0)=\tilde\alpha^{(i)}_{j'}$ are also
independent of $\xi_0$,  and  (\ref{nalpha}) reduces to
$\tilde\Phi_i(\mbox{\boldmath $\xi$})=\sum_{j'=1}^{j'_{\max}}
\tilde\alpha^{(i)}_{j'}\Phi^{S(A)}_{j'}(\mbox{\boldmath $\xi$})$.\\
(O) If  $V^{pair}(x_{ij})=V^{hosc}(x_{ij})$ and $U^{pair}_{ij'}=0$, then
$\tilde\epsilon_i =E^{S(A)}_i$ and $\tilde\alpha^{(i)}_{j'}=\delta_{ij'}$.

For the short-range  barrier potentials $V (\xi_{0}, x_{i}({\mbox{\boldmath $\xi$}}))$
in terms of the asymptotic cluster functions
$\tilde\Phi_{j}({\mbox{\boldmath $\xi$}})\rightarrow\tilde\Phi_{j}(\xi_{0},
{\mbox{\boldmath $\xi$}})$ at $|\xi_0|\rightarrow \infty$ the asymptotic boundary conditions
for the solution $\Psi(\xi_0,{\mbox{\boldmath $\xi$}})=\{\Psi_{i_{o}}(\xi_0,{\mbox{\boldmath $\xi$}})\}_{i_{o}=1}^{N_o}$
in the asymptotic region $|{\mbox{\boldmath $\xi$}}|/|\xi_0|\ll 1$ have the form  \cite{casc11}
 \begin{eqnarray}
 &&\hspace{-0.2cm}\Psi_{i_o}^{\stackrel{\scriptstyle \gets}{\scriptstyle\to}}
 (\xi_0\to\pm\infty,{\mbox{\boldmath $\xi$}})\rightarrow \tilde\Phi_{i_o}({\mbox{\boldmath $\xi$}})
 \frac{\exp\left(\mp\imath \left(p_{i_o}\xi_0\right)\right)}
 {\sqrt{p_{i_o}}} +
 \sum_{j=1}^{N_o}\tilde\Phi_{j}({\mbox{\boldmath $\xi$}})\frac{\exp\left(\pm
 \imath \left(p_{j}\xi_0\right)\right)}{\sqrt{p_{j}}}R_{ji_o}^{\stackrel{\scriptstyle \gets}{\scriptstyle\to}}(E), \nonumber\\
 &&\hspace{-0.2cm}\Psi_{i_o}^{\stackrel{\scriptstyle \gets}{\scriptstyle\to}}(\xi_0\to\mp\infty,{\mbox{\boldmath $\xi$}})\rightarrow\sum_{j=1}^{N_o}\tilde\Phi_{j}({\mbox{\boldmath $\xi$}})
 \frac{\exp\left(\mp\imath \left(p_{j}\xi_0\right)\right)}{\sqrt{p_{j}}} T_{ji_o}^{\stackrel{\scriptstyle \gets}{\scriptstyle\to}}(E),\label{TR}
 \\
 &&\hspace{-0.2cm}\Psi_{i_o}^{\stackrel{\scriptstyle \gets}{\scriptstyle\to}}(\xi_0,|{\mbox{\boldmath $\xi$}}|\to \infty)\rightarrow 0.\nonumber
\end{eqnarray}
Here $v=\leftarrow,\rightarrow$ indicates the initial direction of
the particle motion along the $\xi_0$ axis, $N_o$ is the number of
open channels at the fixed energy $E$ and momentum
$p^{2}_{i_{o}}=E-E_{i_{o}} >0$ of cluster;
$R_{ji_{o}}^{\leftarrow}=R_{ji_{o}}^{\leftarrow}(E)$,
$R_{ji_{o}}^\rightarrow=R_{ji_{o}}^\rightarrow(E)$ and
$T_{ji_{o}}^{\leftarrow}=T_{ji_{o}}^{\leftarrow}(E)$,
$T_{ji_{o}}^\rightarrow=T_{ji_{o}}^\rightarrow(E)$ are the unknown
amplitudes of the
reflected and transmitted waves. We can rewrite Eqs. (\ref{TR})
in the matrix form
$\mbox{\boldmath$\Psi$}=\mbox{\boldmath$\tilde\Phi$}^T\mbox{\boldmath$F$}$
describing the incident wave and the outgoing waves at
$\xi_0^{+}\to + \infty$ and $\xi_0^{-}\to - \infty$ as
 \begin{eqnarray}
 &&\hspace{-0.9cm}\left(
 \begin{array}{ll}
 \mbox{\boldmath$F$}_\rightarrow(\xi_0^+) & \mbox{\boldmath$F$}_\leftarrow(\xi_0^+) \\
 \mbox{\boldmath$F$}_\rightarrow(\xi_0^-) & \mbox{\boldmath$F$}_\leftarrow(\xi_0^-)
 \end{array} \right)\!=\!\left(
 \begin{array}{ll}
 \mathbf{0}             & \mathbf{ X}^{(-)}(\xi_0^+)\\
 \mathbf{ X}^{(+)}(\xi_0^-) & \mathbf{0}
 \end{array} \right)\!+\!\left(
 \begin{array}{ll}
 \mathbf{0}             & \mathbf{ X}^{(+)}(\xi_0^+)\\
 \mathbf{ X}^{(-)}(\xi_0^-) & \mathbf{0}
 \end{array} \right)\mathbf{S}.
 \end{eqnarray}
Here the unitary and symmetric scattering matrix $\mathbf{S}$
 \begin{eqnarray} \mathbf{S}=\left(
 \begin{array}{ll}
 \mathbf{ R}_\rightarrow & \mathbf{ T}_\leftarrow\\
 \mathbf{ T}_\rightarrow & \mathbf{ R}_\leftarrow
 \end{array} \right),\quad \mathbf{S}^\dag\mathbf{S}=\mathbf{S}\mathbf{S}^\dag=\mathbf{I},\label{new3344a}
 \end{eqnarray}
where $\mathbf{S}^\dag$ is the conjugate transpose of $\mathbf{S}$. It is composed of the matrices,
whose elements are reflection and transmission amplitudes that enter Eqs. (\ref{TR})
and possess the following properties\cite{casc11,stankin11}:
 \begin{eqnarray}
 &&\mathbf{ T}_\rightarrow^{\dag}\mathbf{ T}_\rightarrow+\mathbf{
 R}_\rightarrow^{\dag}\mathbf{R}_\rightarrow=\mathbf{I}_{oo}=\mathbf{ T}_\leftarrow^{\dag}\mathbf{ T}_\leftarrow+\mathbf{
 R}_\leftarrow^{\dag}\mathbf{R}_\leftarrow,\nonumber\\
  &&\mathbf{ T}_\rightarrow^{\dag}\mathbf{ R}_\leftarrow+\mathbf{ R}_\rightarrow^{\dag}\mathbf{
 T}_\leftarrow=\mathbf{0}=\mathbf{ R}_\leftarrow^{\dag}\mathbf{ T}_\rightarrow+\mathbf{ T}_\leftarrow^{\dag}\mathbf{
 R}_\rightarrow,\label{new32}\\
 && \mathbf{ T}_\rightarrow^T=\mathbf{ T}_\leftarrow,\quad \mathbf{ R}_\rightarrow^T=\mathbf{
 R}_\rightarrow,\quad \mathbf{ R}_\leftarrow^T=\mathbf{
 R}_\leftarrow.\nonumber
 \end{eqnarray}

 \begin{figure}[t]
\begin{center} \epsfig{file=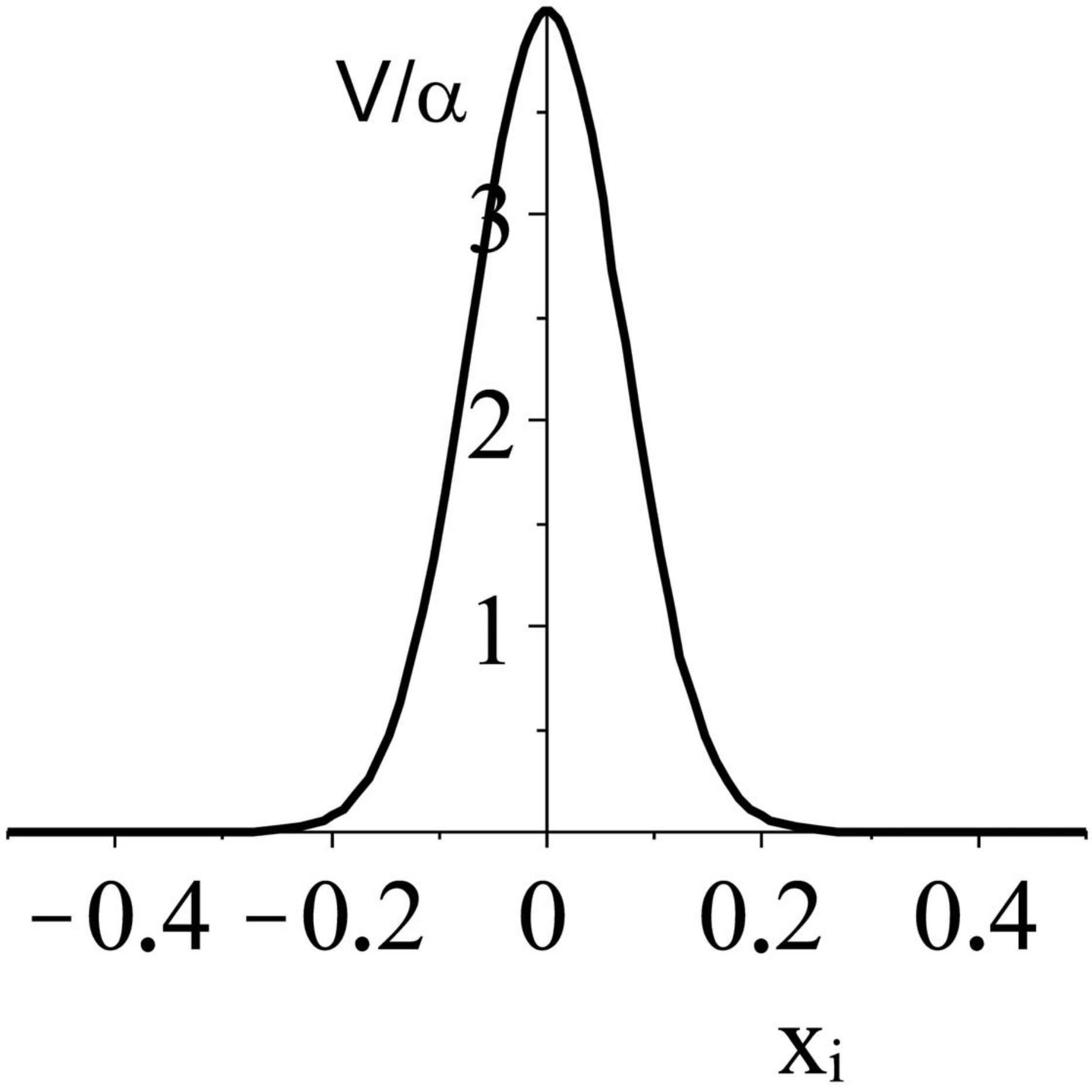,width=0.32\textwidth,angle=0 }\hfil
 \epsfig{file=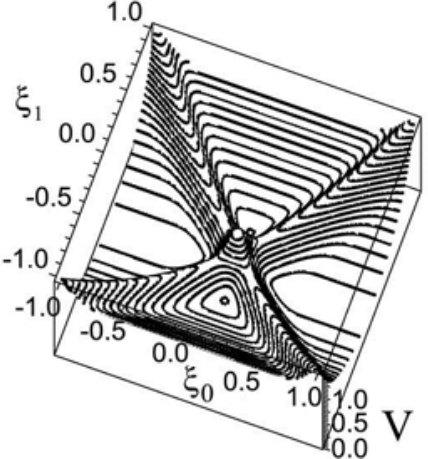,width=0.32\textwidth,angle=0 }
\end{center}
\caption{The Gaussian-type potential
  (\ref{mo3}) at $\sigma=0.1$ (in oscillator units) and the corresponding 2D barrier potential at $\alpha=1/10$,  $\sigma=0.1$ }
\label{ga}
\end{figure}

\begin{figure}[t]
 \epsfig{file=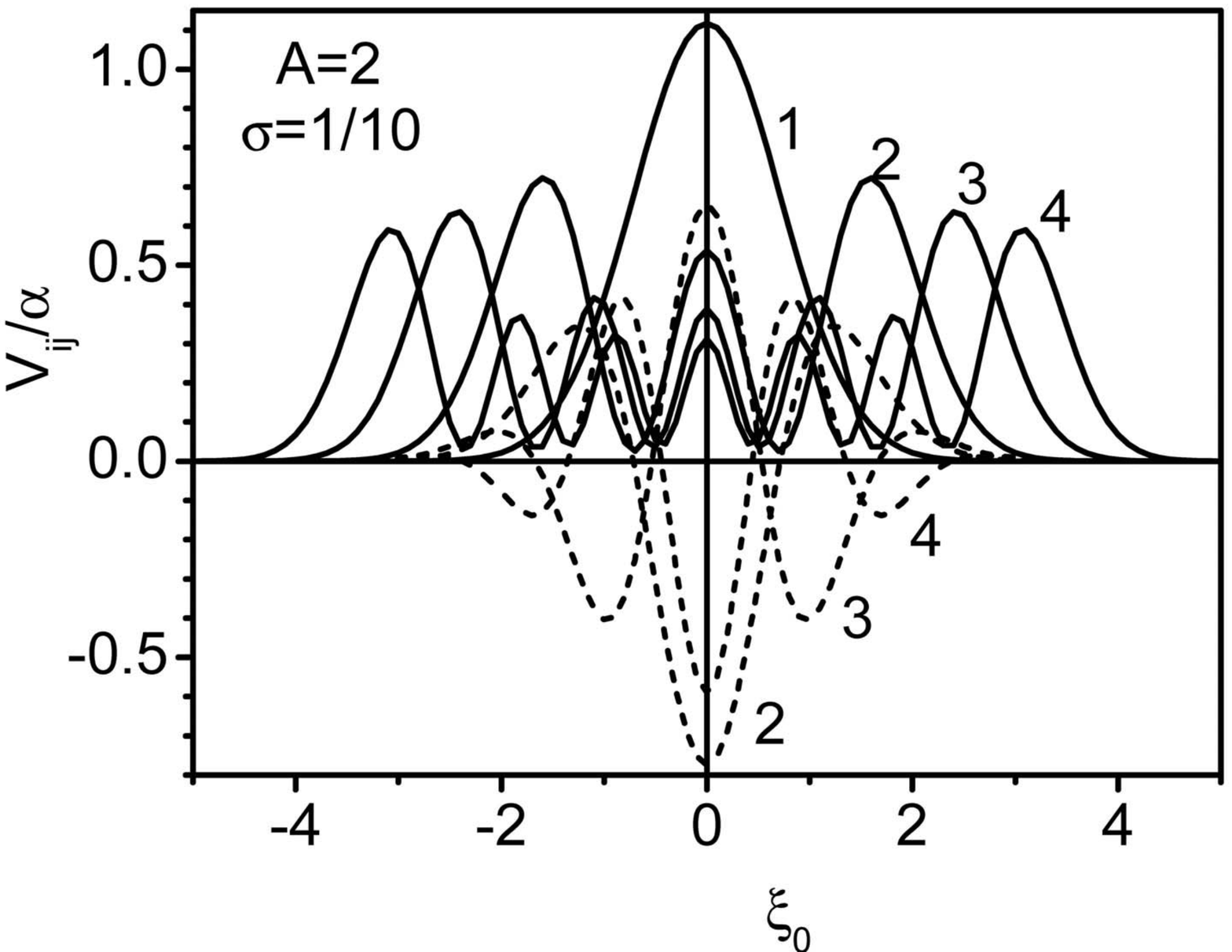,width=0.32\textwidth,angle=0}
 \epsfig{file=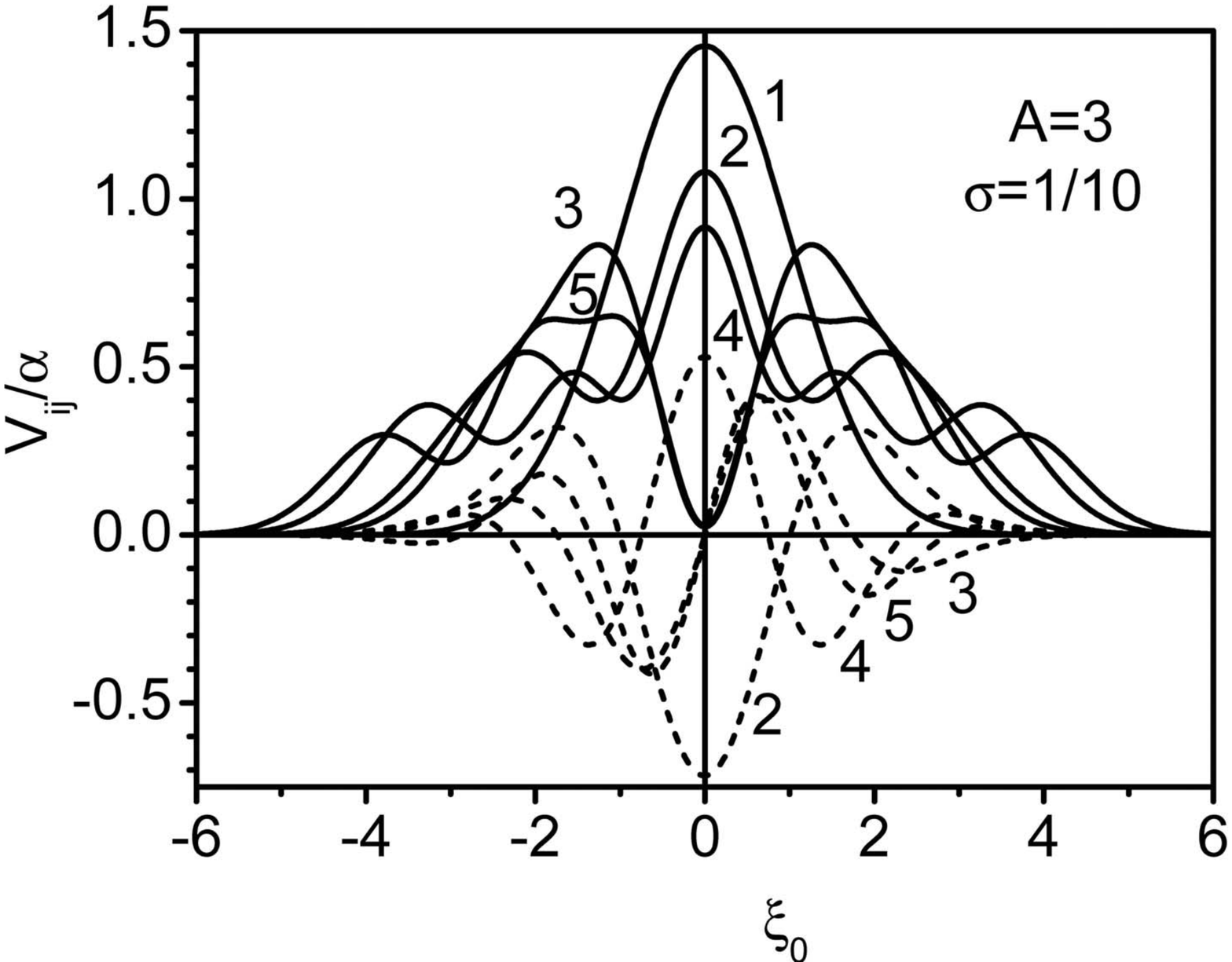,width=0.32\textwidth,angle=0}
 \epsfig{file=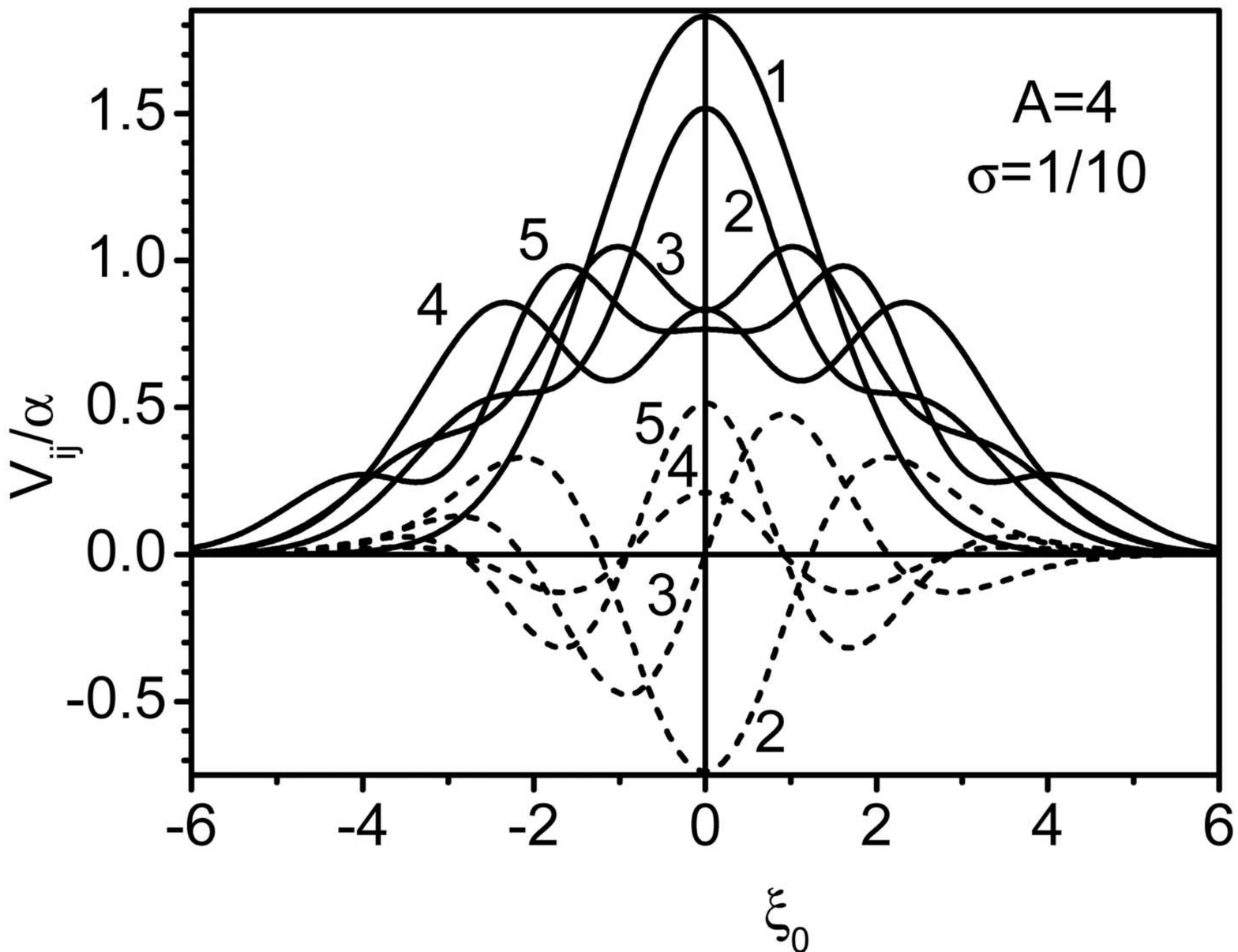,width=0.32\textwidth,angle=0}\\
 \epsfig{file=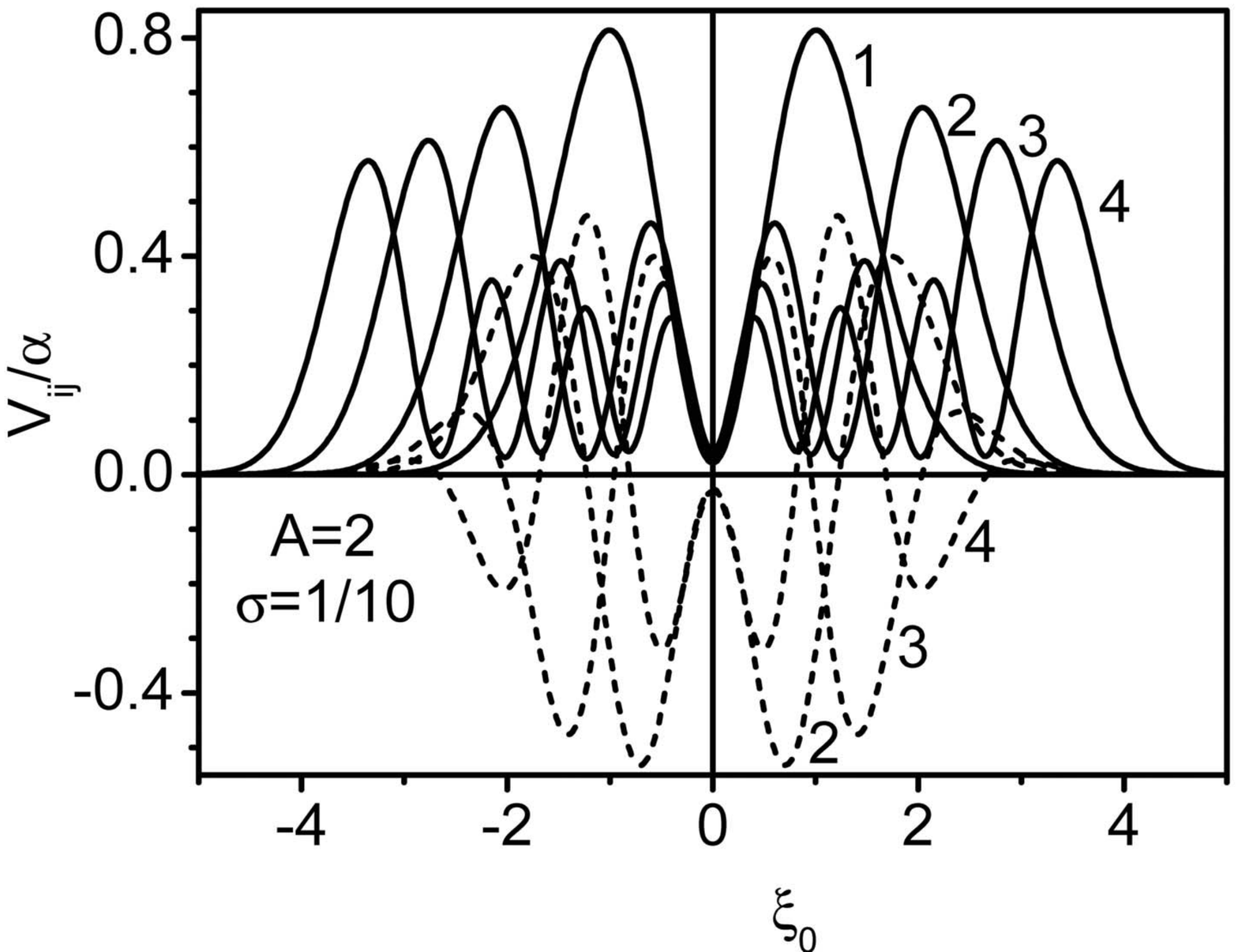,width=0.32\textwidth,angle=0}
 \epsfig{file=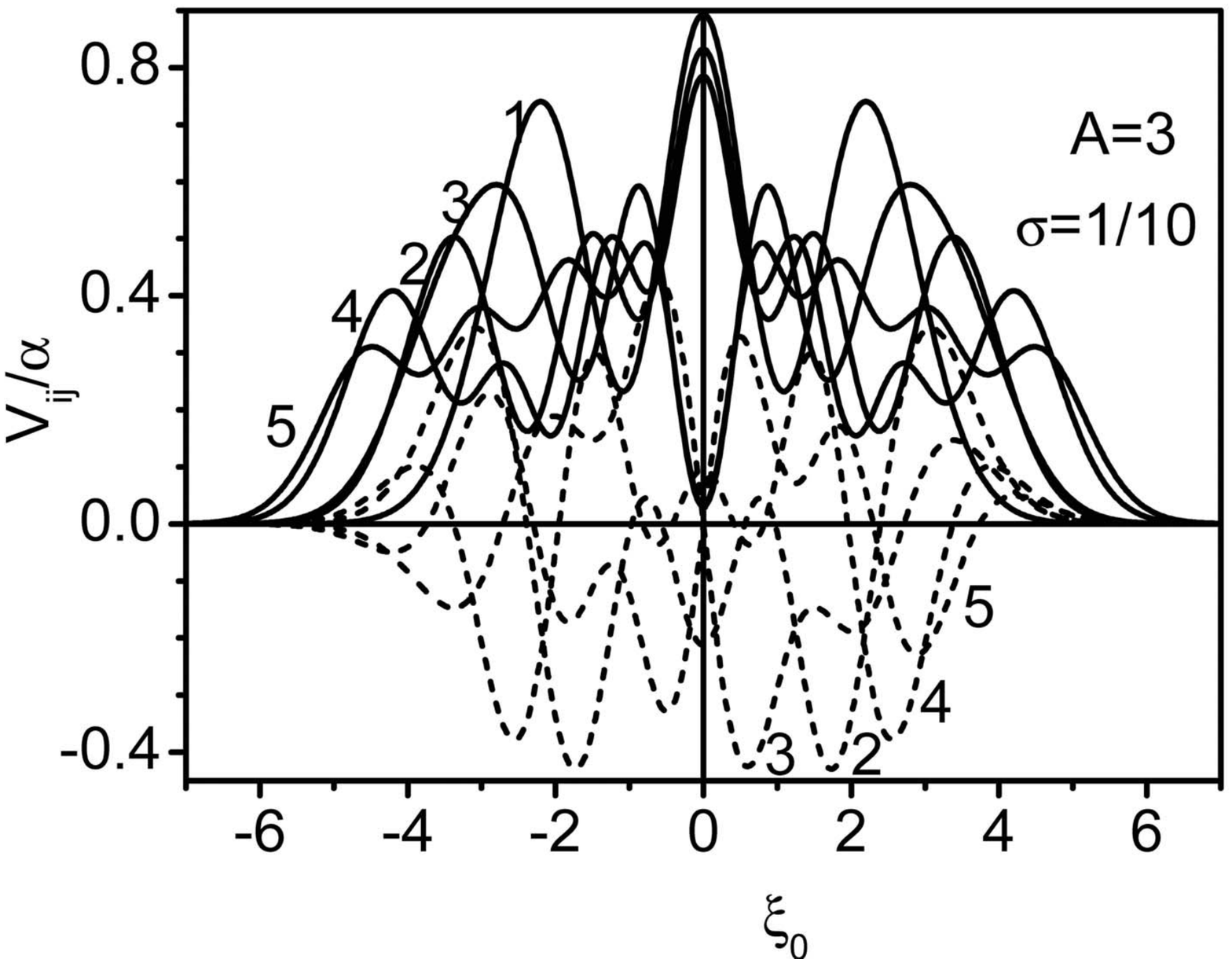,width=0.32\textwidth,angle=0}
 \epsfig{file=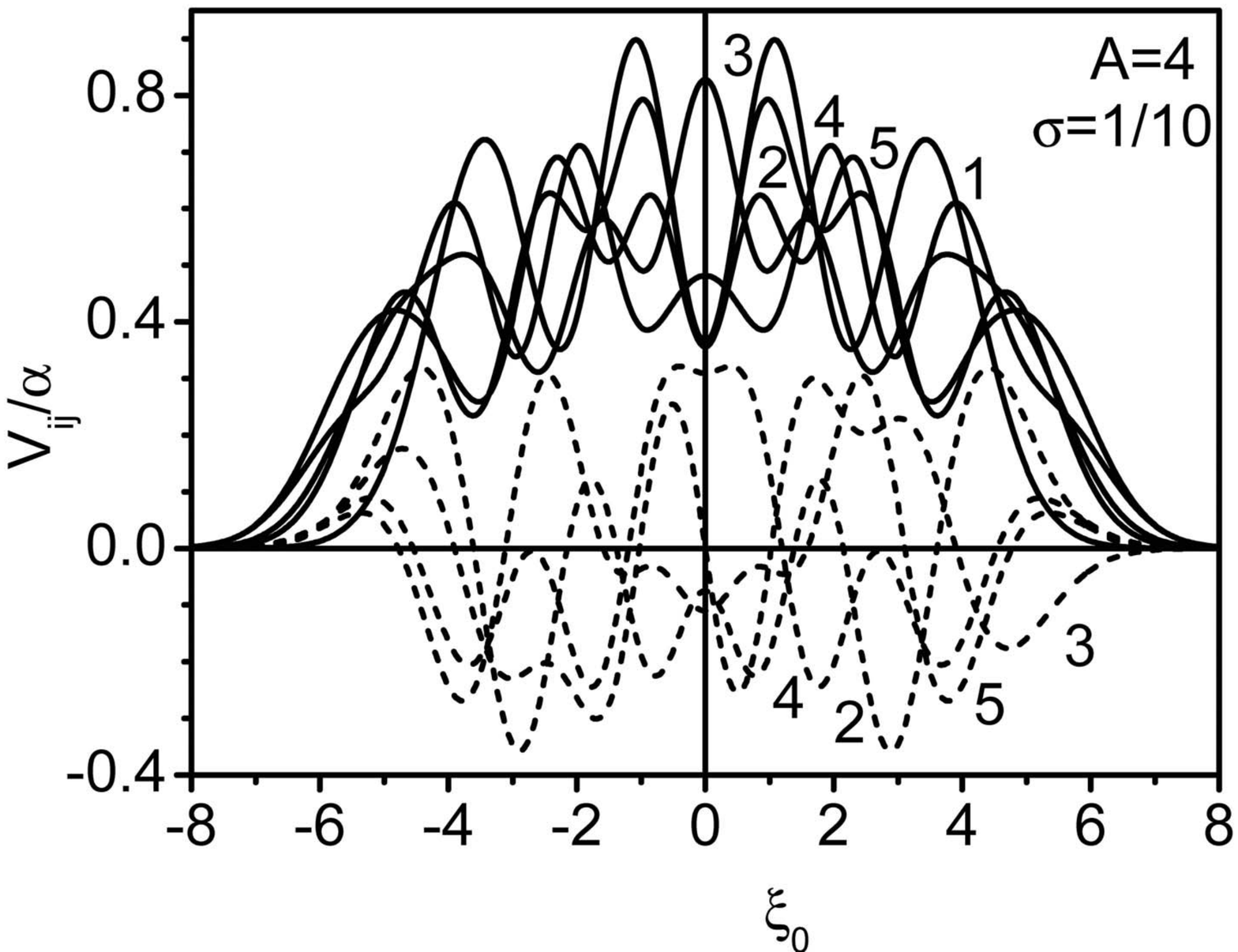,width=0.32\textwidth,angle=0}
\caption{Diagonal $V_{jj}$ (solid lines) and nondiagonal $V_{j1}$,
 (dashed lines) effective potentials for $A=2$,  $A=3$ and $A=4$ of the  S- (upper panels) and A- (lower panels) of the particles
 at $\sigma=1/10$
 } \label{vij}
\end{figure}

\section{Close-coupling Equations in the SCR}

We seek for the solution of problem (\ref{mo5}) in the symmetrized
coordinates in the form of  Galerkin (G)  expansion over the
asymptotic cluster functions $\tilde\Phi_j(\mbox{\boldmath
$\xi$})$  corresponding to the eigenvalues $\tilde \epsilon_i$,
which are also independent of $\xi_0$, from (\ref{noparam}) under
the (G) condition, with unknown coefficient functions
$\chi_{ji_{o}}(\xi_0)$:
\begin{eqnarray}
\Psi_{i_{o}}(\xi_0,\mbox{\boldmath $\xi$})=\sum_{j=1}^{j_{\max}}\tilde\Phi_j(\mbox{\boldmath $\xi$})\chi_{ji_{o}}(\xi_0),\,\,\chi_{ji_{o}}(\xi_0)=\int d^{A-1}\mbox{\boldmath $\xi$} \tilde\Phi_j(\mbox{\boldmath $\xi$})\Psi_{i_{o}}(\xi_0,\mbox{\boldmath $\xi$}).
\label{dec}
\end{eqnarray}

The set of  close-coupling Galerkin equations in the symmetrized coordinates has the form
\begin{eqnarray}\label{mo3e}
\left[-\frac{d^2}{d \xi_0^2}
+\tilde \epsilon_i-E\right]\chi_{ii_o}(\xi_0)
+\sum_{j=1}^{j_{\max}} \tilde V_{ij}(\xi_0) \chi_{ji_o}(\xi_0) =0,\qquad\qquad
\\ \nonumber
\end{eqnarray}
where
the effective potentials  $\tilde V_{ij}(\xi_0)$
are calculated using
the set of  eigenvectors $\tilde\alpha^{(i)}_{j'}$ of
the noparametric
algebraic problem (\ref{noparam})
under the above condition (G): $U_{ij'}(\xi_0)=U^{pair}_{ij'}\neq0$,
\begin{eqnarray}\label{sumint}
\!\!\!\!\!\!\!\!\!\!\!\!\!\!\tilde V_{ij}(\xi_0)=
\sum_{j'=1}^{j'_{\max}}\sum_{j''=1}^{j'_{\max}}\tilde\alpha^{(i)}_{j'}
V_{j'j''}(\xi_0)\tilde\alpha^{(j)}_{j''},
\end{eqnarray}
and the integrals $V_{ij'}(\xi_0)$ are defined in
(\ref{integrals}) and calculated in CAS MAPLE. In the examples
considered below, we put $U_{ij'}(\xi_0)=U^{pair}_{ij'}=0$ in
(\ref{noparam}), then we have the (O) condition: $\tilde\epsilon_i
=E_i^{S(A)}$, $\tilde\alpha^{(i)}_{j'}=\delta_{ij'}$ and $\tilde
V_{ij}(\xi_0)=V_{ij}(\xi_0)$. The repulsive barrier is chosen to
have the Gaussian shape
\begin{eqnarray}
V(x_i)=\frac{\alpha}{\sqrt{2\pi}\sigma}\exp(-\frac{x_i^2}{\sigma^2}).\label{mo3}
\end{eqnarray}
Figure \ref{ga} illustrates the Gaussian potential and the
corresponding barrier potentials in the symmetrized coordinates at
$A=2$. This potential has the oscillator-type shape, and two
barriers are crossing at the right angle. In the case $A\ge 3$,
the hyperplanes of barriers are crossing at the right angle, too.

The effective potentials $V_{ij}(\xi_0)$
calculated using the
\textit{algorithm SCR} \cite{cascnedve1}
and
\textit{algorithm DC}
(see Section 5),
 are shown in Fig. \ref{vij}. In comparison with the symmetric basis, for antisymmetric one the
increase of the numbers $i$ and/or $j$ results in stronger oscillation of the effective potentials
 $V_{ij}$ and weaker decrease of them to zero at $\xi_0\to\infty$.
At $A=2$, all effective potentials are even functions, and at
$A\geq 3$, some effective potentials are odd functions.

Thus, the scattering problem (\ref{mo5}) with the asymptotic
boundary conditions (\ref{TR}) is reduced to the boundary-value
problem for the set of close-coupling equations in the Galerkin
form (\ref{mo3e}) under the boundary conditions at $d=1$,
$\xi_0=\xi_{\min}$ and $\xi_0=\xi_{\max}$:
 \begin{eqnarray}
 \frac{d \mbox{\boldmath$F$}(\xi_0)}{d \xi_0}\biggl|_{\xi_0=\xi_{\min}}\!\!\!\!\!=
 \mathbf{\mathcal{R}}(\xi_{\min})\mbox{\boldmath$F$}(\xi_{\min}),\,
 \frac{d \mbox{\boldmath$F$}(\xi_0)}{d \xi_0}\biggl|_{\xi_0=\xi_{\max}}\!\!\!\!\!=
 \mathbf{\mathcal{R}}(\xi_{\max})\mbox{\boldmath$F$}(\xi_{\max}),
 \label{new12}
 \end{eqnarray}
where $\mathbf{\mathcal{R}}(\xi)$ is an unknown $j_{\max}\times j_{\max}$
matrix function,
$\mbox{\boldmath$F$}(\xi_0)=\{\mbox{\boldmath$\chi$}_{i_o}(\xi_0)\}_{i_o=1}^{N_{o}} =
\{\{\chi_{ji_o}(\xi_0)\}_{j=1}^{j_{\max}}\}_{i_o=1}^{N_{o}}$
is the required $j_{\max}\times N_o$ matrix solution, and $N_o$ is the
number of open channels, $N_o=\max\limits_{2E\geq \tilde \epsilon_j}j\leq j_{\max}$,
calculated using the third version of KANTBP 3.0 program \cite{kantbp2,kantbp3},
implemented in CAS MAPLE and  FORTRAN environment and described in \cite{stankin11,casc11}.

\begin{figure}[t]
\epsfig{file=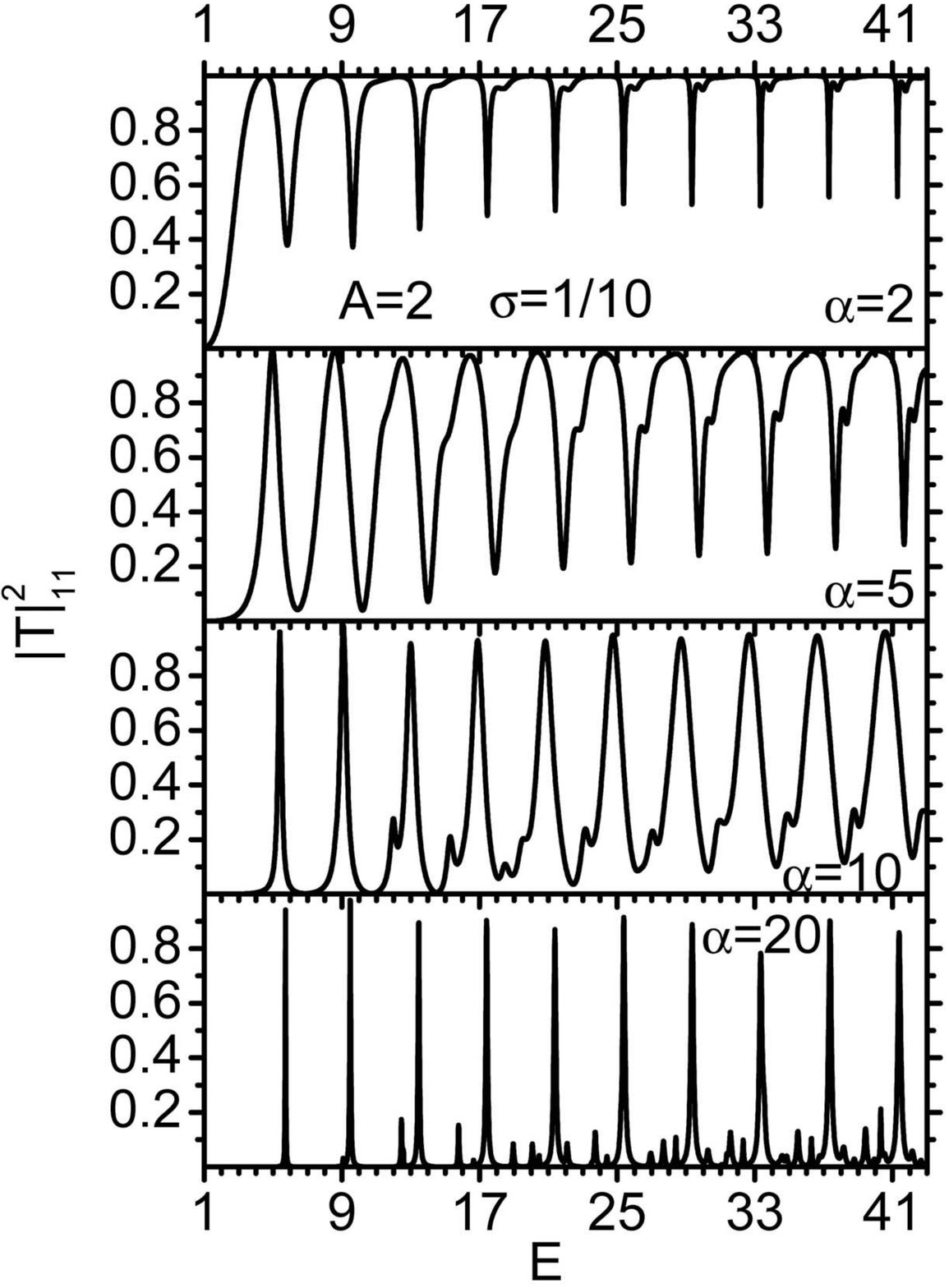,width=0.31\textwidth}
\epsfig{file=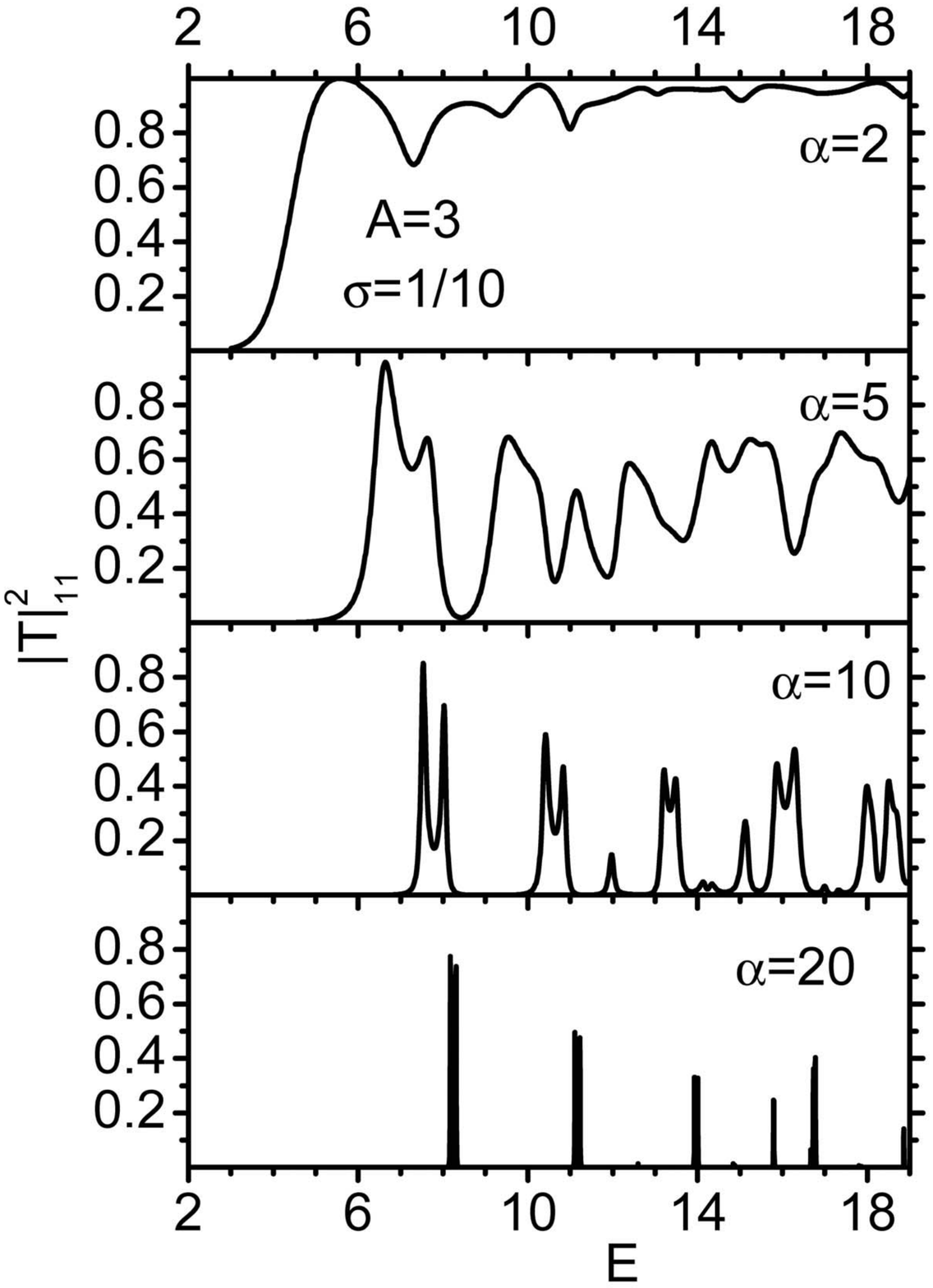,width=0.31\textwidth}
\epsfig{file=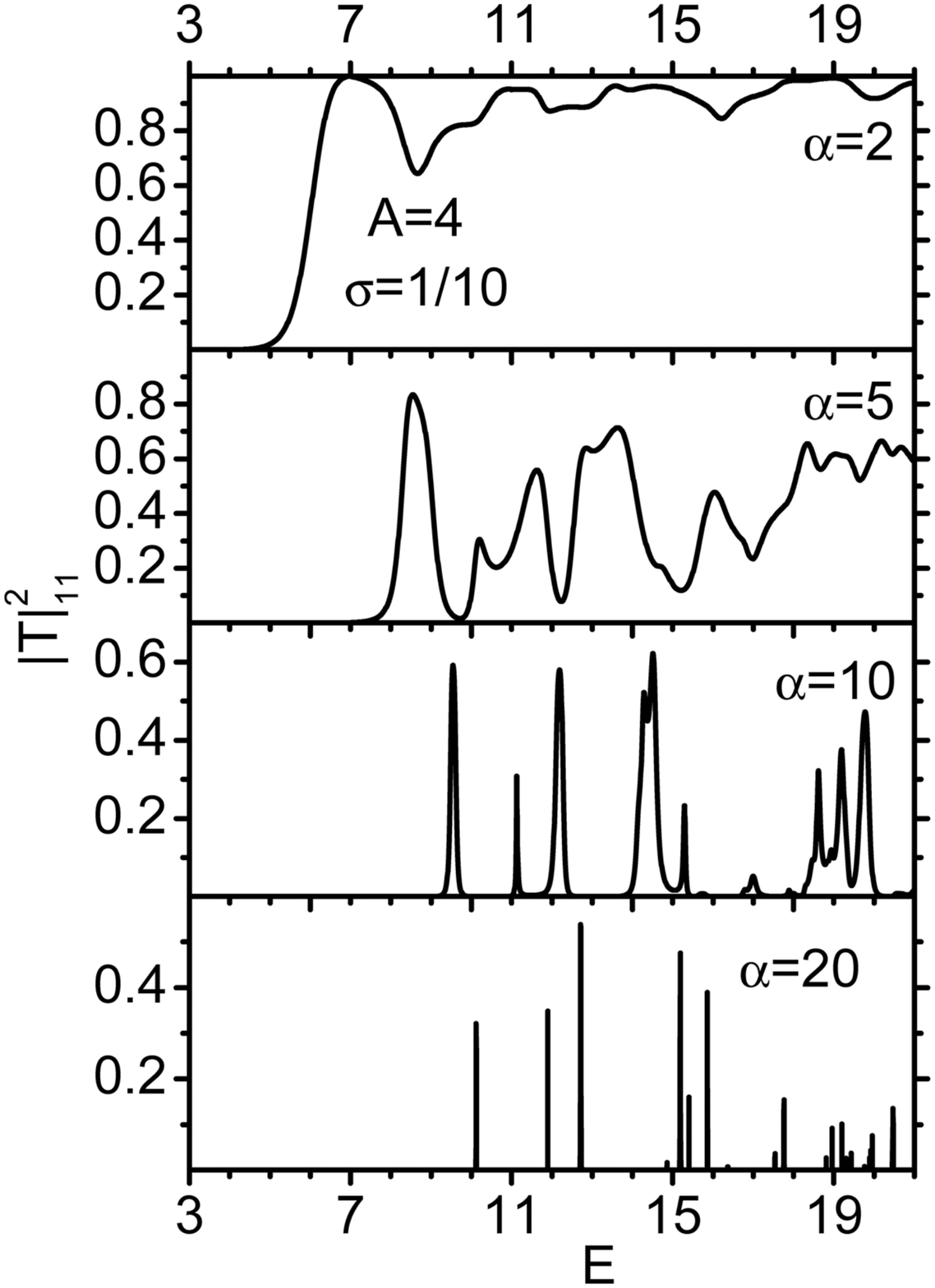,width=0.31\textwidth}\\
\epsfig{file=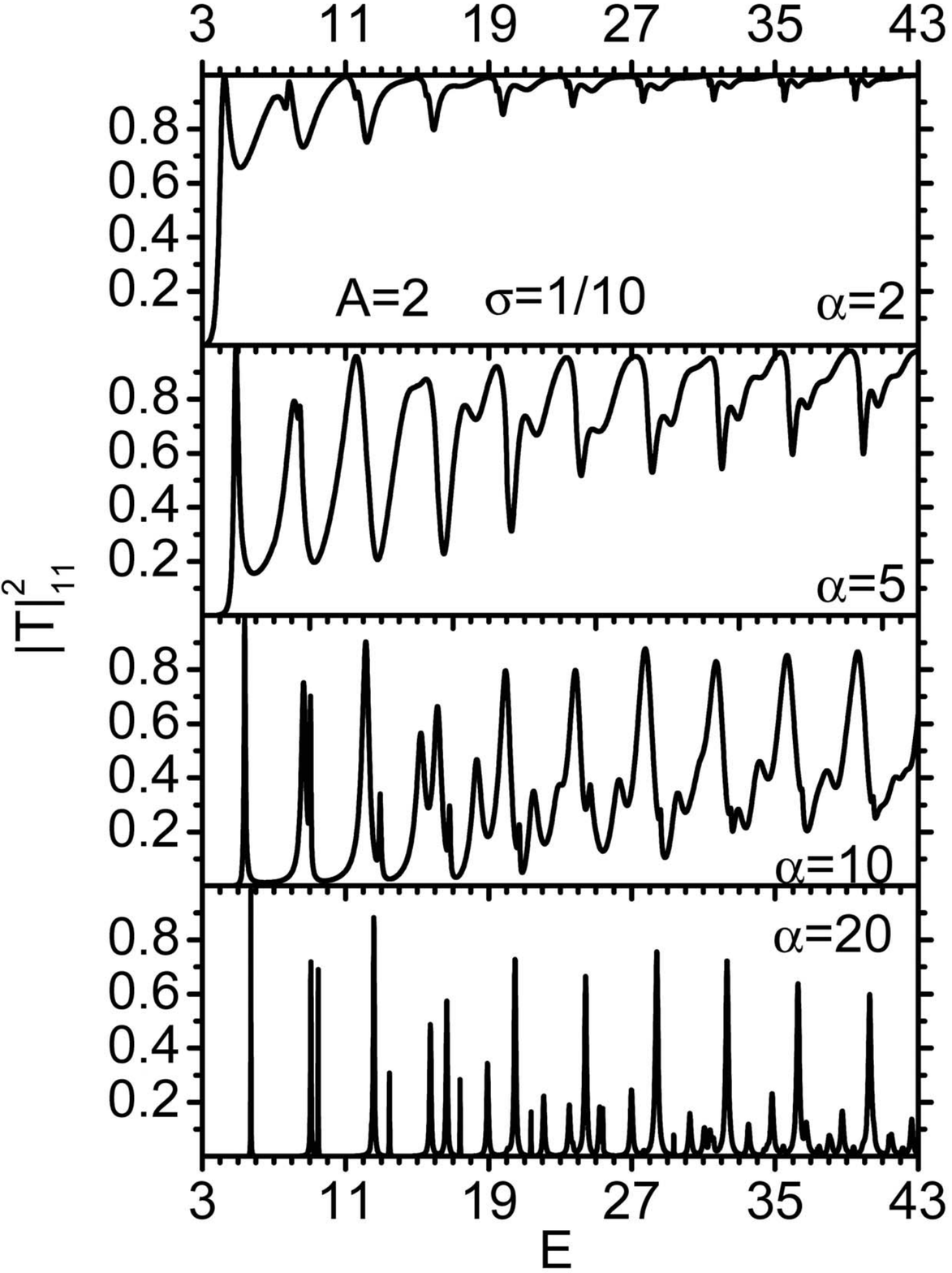,width=0.31\textwidth}
\epsfig{file=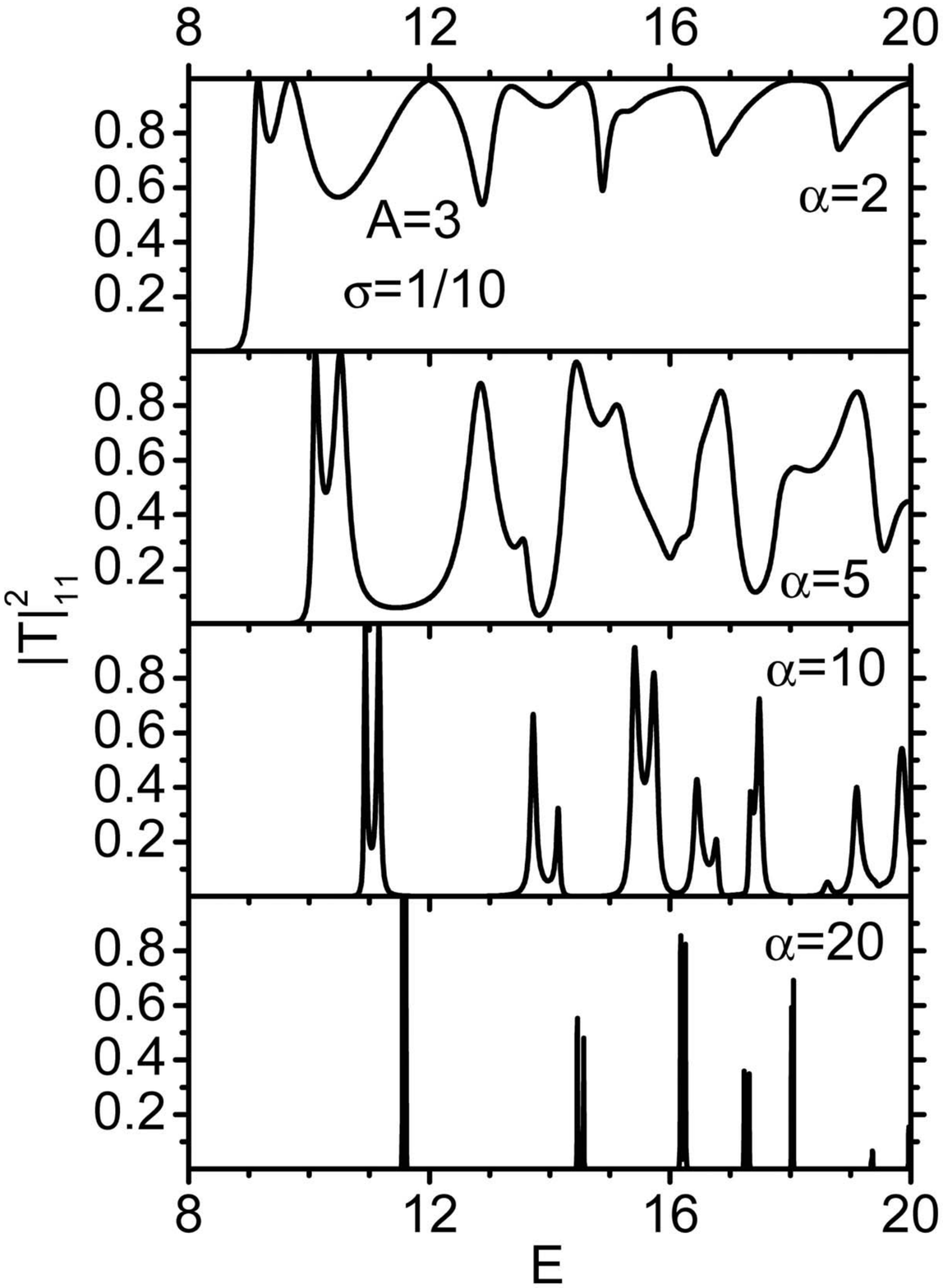,width=0.31\textwidth}
\epsfig{file=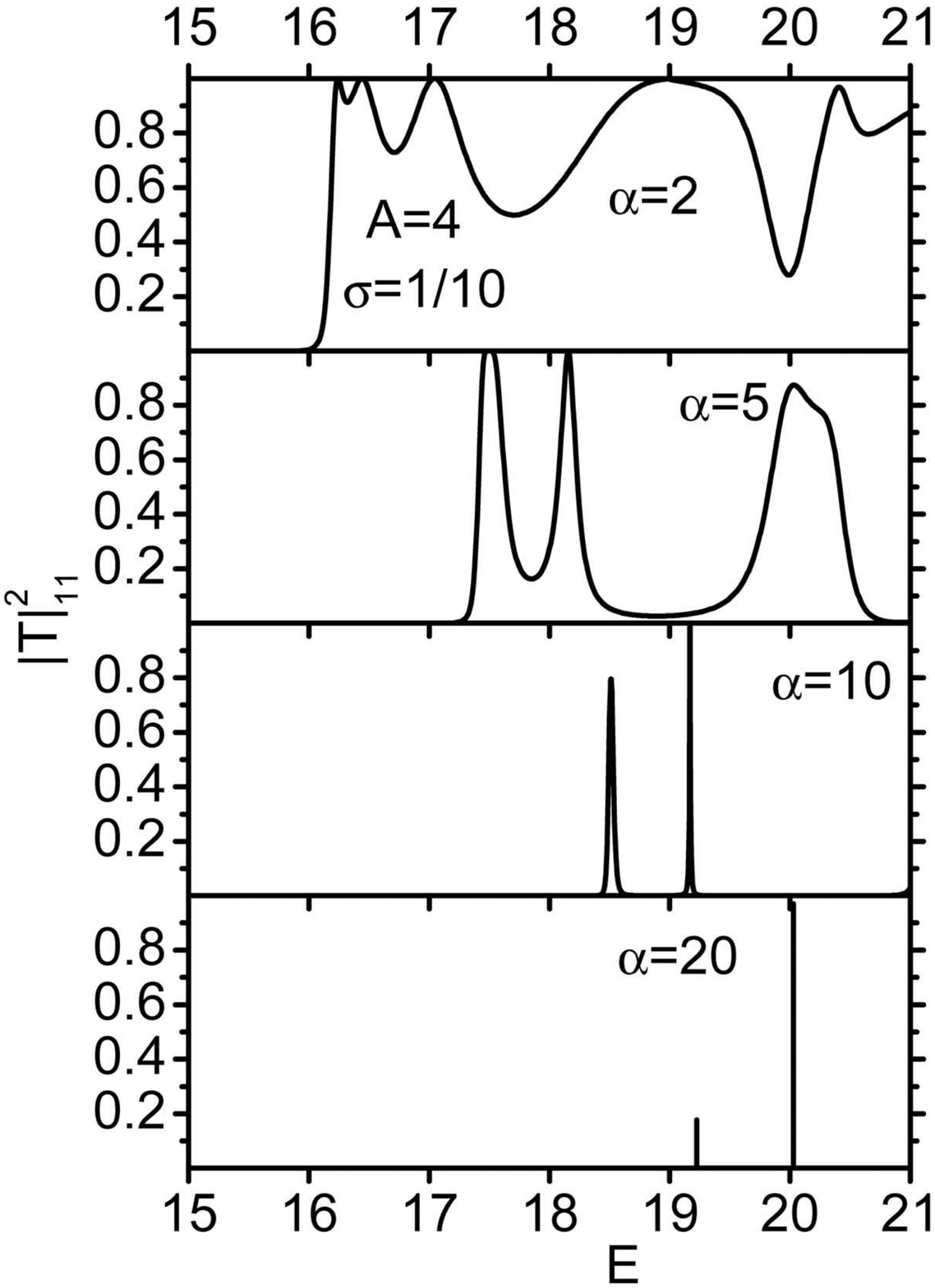,width=0.31\textwidth}
\caption{The total transmission probability $|T|^2_{11}$ vs energy
$E$ (in oscillator units) for the system of $A=2,3,4$  S- (upper
panels) and A- (lower panels)  particles coupled by the oscillator
potential and being initially in the ground cluster state
penetrating through the repulsive  Gaussian-type potential
barriers (\ref{mo3}) with $\sigma=0.1$ and $\alpha=2,5,10,20$
 } \label{nt11}
\end{figure}

\begin{figure}[t]
\epsfig{file=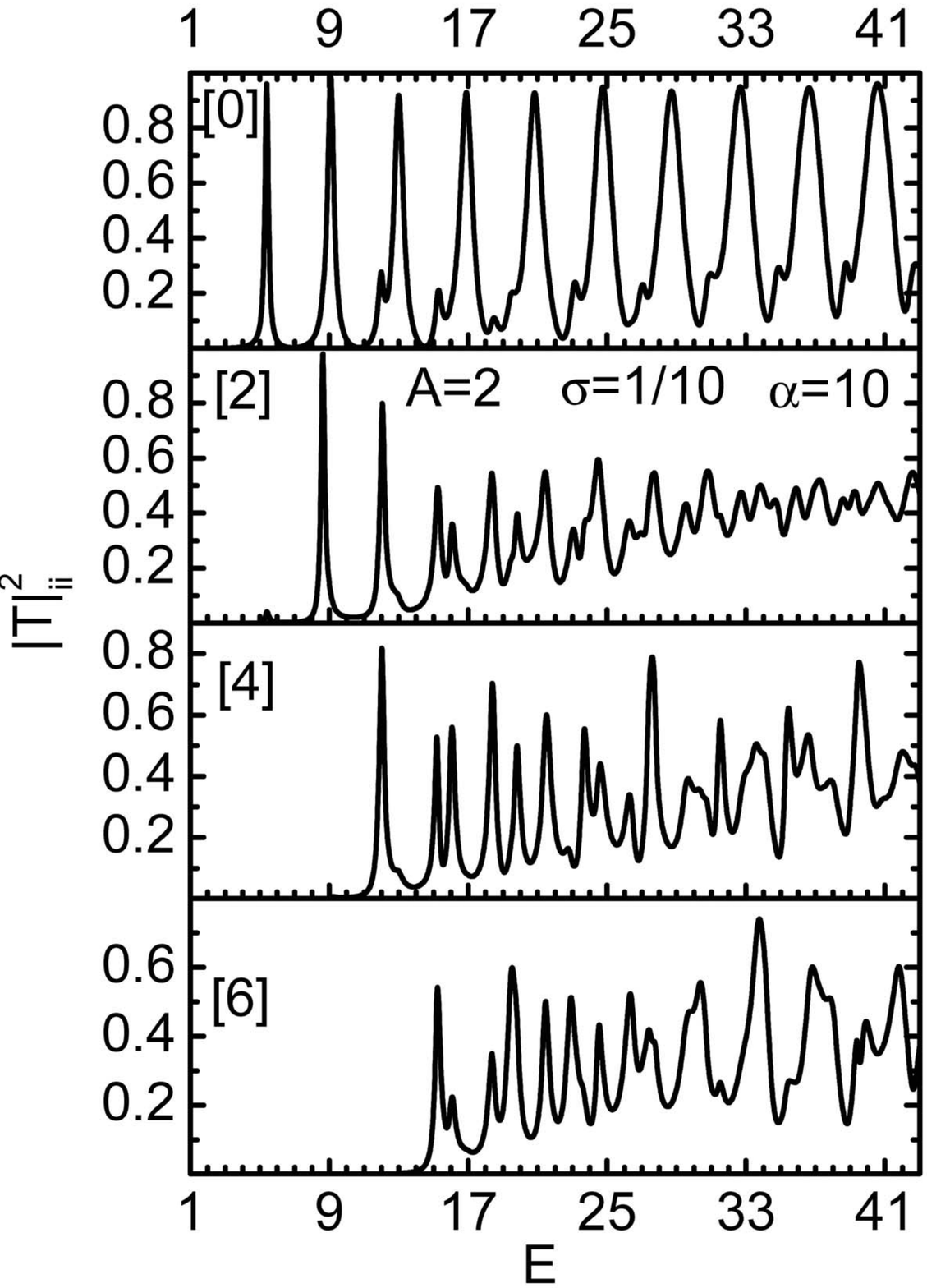,width=0.31\textwidth}
\epsfig{file=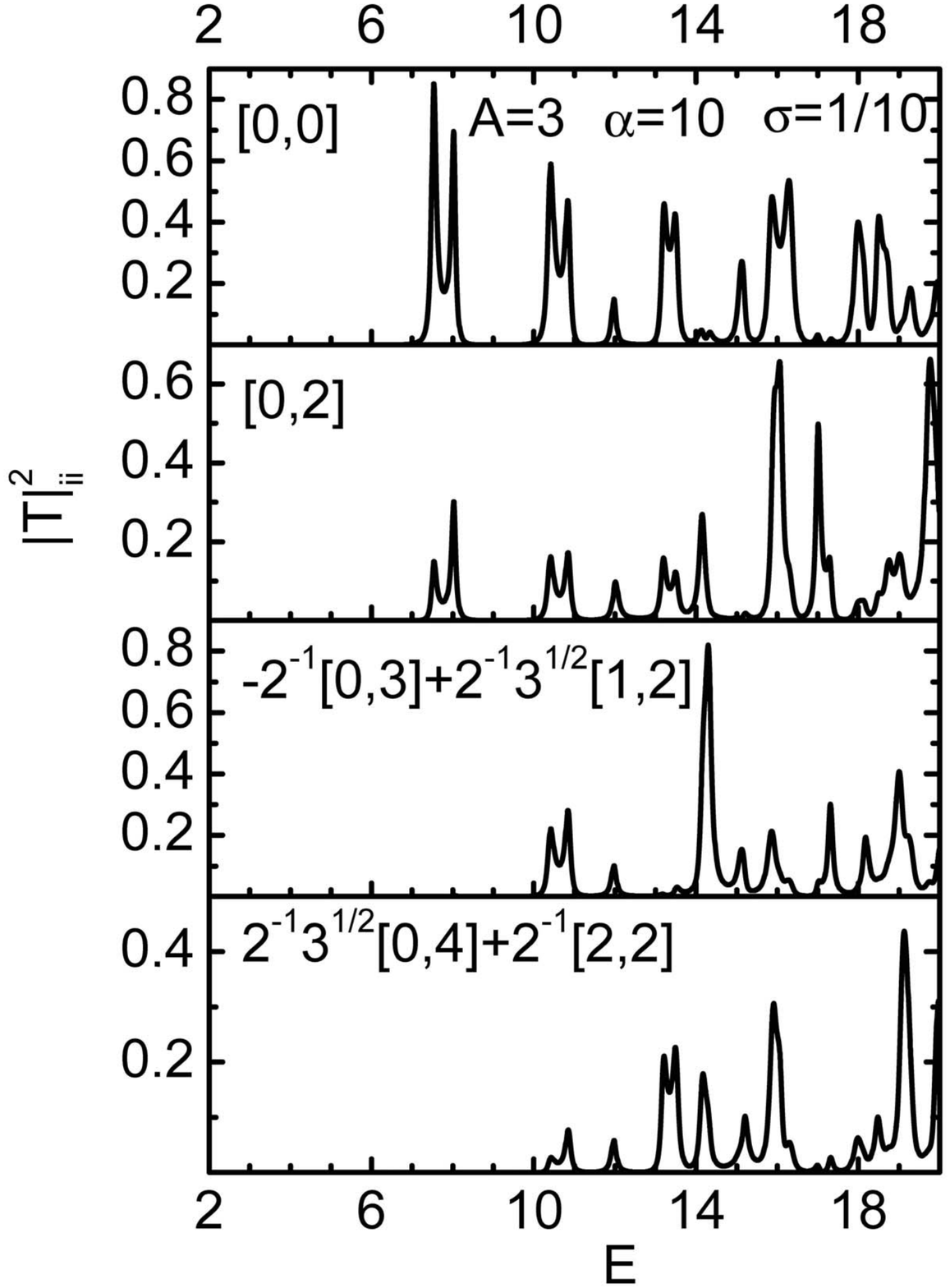,width=0.31\textwidth}
\epsfig{file=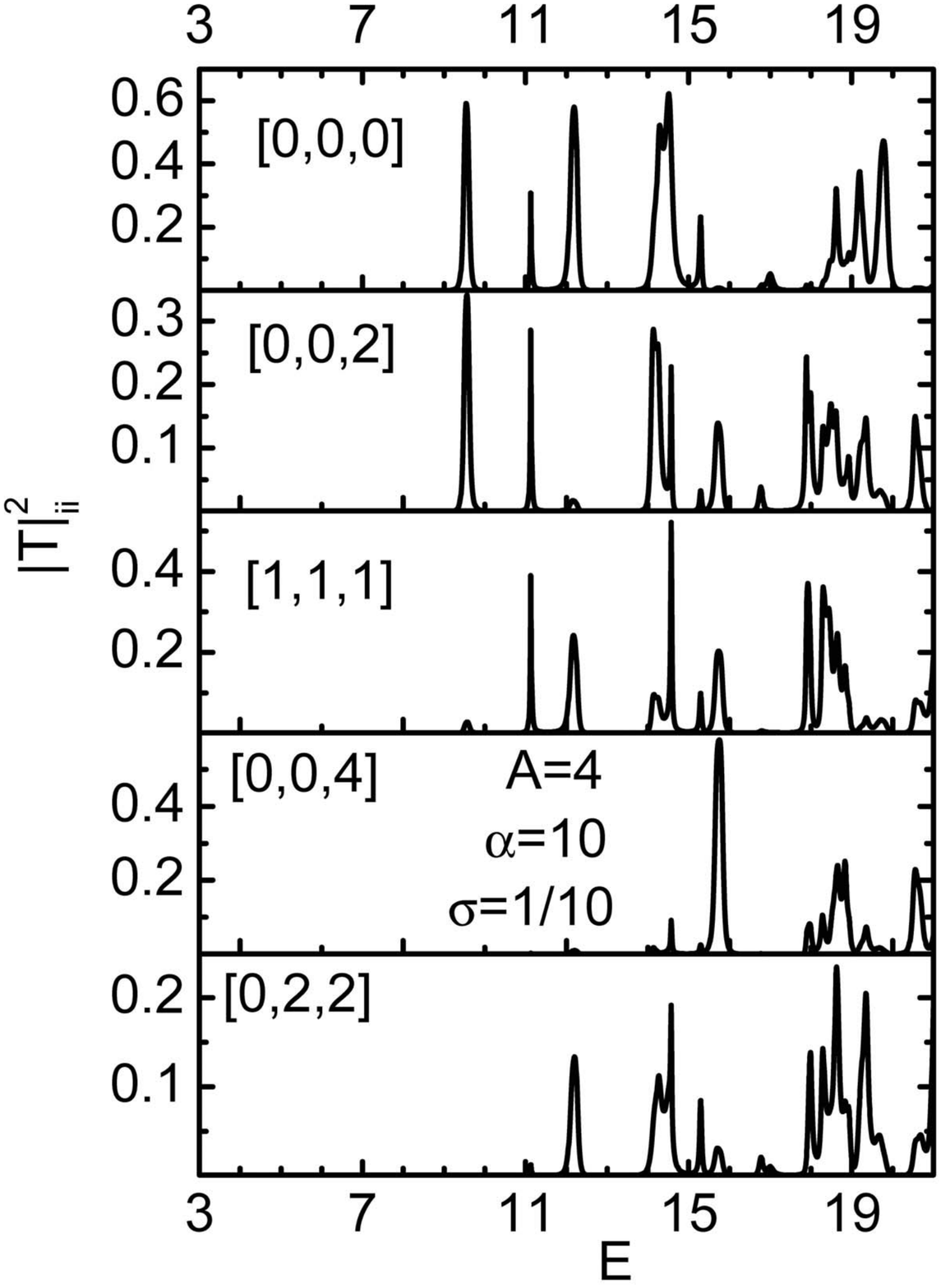,width=0.31\textwidth}
\caption{The total transmission probability $|T|^2_{ii}$ vs the energy $E$ (in oscillator units) for the system of $A=2,3,4$ particles,
 coupled by the oscillator potential and being initially in the ground and excited S-states, penetrating
through the repulsive  Gaussian-type potential barriers (\ref{mo3})
 with $\sigma=0.1$ and $\alpha=10$.
We use the notation of the S-states,
$[i_1,...,i_{A-1}]=1/\sqrt{N_\beta}\sum_{i_1',...,i_{A-1}'}
\prod\bar\Phi_{i_k'}(\xi_k)$, with summation over all ($N_\beta$)
multiset permutations of $i_1,...,i_{A-1}$ of $A-1$-dimensional
oscillator functions \cite{cascnedve1}
 }
 \label{nt10}
\end{figure}
\begin{table}[t]
\caption{
Resonance values of the energy $E_S$ ($E_A$) for S (A) states for $A=2,3,4$
($\sigma=1/10$, $\alpha=20$)
  with approximate eigenvalues
$E_{i}^{D}$, for the first ten states $i=1,...,10$, calculated
using the truncated oscillator basis (D) till
$j_{\max}=136,816,1820$ at $A=2,3,4$. The asterisk labels  two
overlapping peaks of transmission probability } \label{res2}
\begin{tabular}{|r||l|l|l|l|l|l|l|l|l|l|} \hline
$i$ &1 &2 &3 &4 &5 &6 &7 &8 &9 & 10 \\ \hline
\multicolumn{11}{|c|}{$A=2$}\\ \hline
$E_S$   &5.72 &9.06 &9.48 &12.46 &12.57 &13.46 &15.74 &15.78 &16.65 & 17.41 \\ \hline
$E_A$   &5.71 &9.06 &9.48 &12.45 &12.57 &13.45 &15.76$^*$ &15.76$^*$ &16.66 &17.40 \\ \hline
$E^{D}_i$ &5.76 &9.12 &9.53 &12.52 &12.64 &13.52 &15.81 &15.84 &16.73 &17.47  \\ \hline\hline
\multicolumn{11}{|c|}{$A=3$}\\\hline
$E_S$   &8.18 &11.11 &     &12.60 &13.93&     &14.84 &15.79&  &16.67\\\cline{2-11}
        &8.31 &11.23 &     &      &14.00&     &14.88 &     &  &16.73\\\hline
$E_A$   &     &      &11.55&      &     &14.46&      &     &16.18& \\\cline{2-11}
        &     &      &11.61&      &     &14.56&      &     &16.25&  \\\hline
$E_i^{D}$ &8.19 &11.09 &11.52&12.51 &13.86&14.42&14.74 &15.67&16.11&16.53\\\hline
\multicolumn{11}{|c|}{$A=4$}\\ \hline
$E_S$         &10.12&11.89&12.71&14.86&15.19&15.41&15.86&16.37&17.54&17.76 \\ \hline
$E^{D31}_i$ &10.03&     &12.60&14.71&15.04&     &     &16.18&17.34&17.56  \\ \hline
$E^{D22}_i$ &     &11.76&     &     &     &15.21&15.64&     &     &  \\ \hline
\end{tabular}
\end{table}
\section{Resonance Transmission of a Few Coupled Particles}
In the (O) case, i.e., $V^{pair}(x_{ij})=V^{hosc}(x_{ij})$, the
solution of the scattering  problem described above yields the
reflection and transmission amplitudes $R_{ji_{o}}(E)$ and
$T_{ji_{o}}(E)$ that enter the asymptotic boundary conditions
(\ref{TR}) as unknowns.  $|R_{ji_{o}}(E)|^2$ ($|T_{ji_{o}}(E)|^2$)
is the probability of a transition to the state described by the
reflected (trans\-mit\-ted) wave and, hence, will be referred as
the reflection (transmission) coefficient. Note that
$|R_{ji_{o}}(E)|^2+|T_{ji_{o}}(E)|^2=1$.

In Figs. \ref{nt11} and  \ref{nt10},  we show the energy
dependence of the total transmission probability
$|T|^2_{ii}=\sum_{j=1}^{N_o}|T_{ji}(E)|^2$. This is the
probability of a transition from a chosen  state $i$ into any of
$N_o$ states found from Eq. (\ref{dec}) {by solving the
boundary-value problem in the Galerkin form, (\ref{mo3e}) and
(\ref{new12}),}
 using the KANTBP 3.0 program \cite{kantbp2,kantbp3}
on the finite-element grid
$\Omega_\xi\{-\xi_0^{\max},\xi_0^{\max}\}$ with $N_{\rm elem}$
fourth-order Lagrange elements between the nodes. For S-solutions
at $A=2,3,4$ the following parameters were used:
$j_{\max}=13,21,39$, $\xi_0^{\max}=9.3, 10.5, 12.8$, $N_{\rm
elem}=664, 800, 976$, while for A-solutions we used
 $j_{\max}=13,16,15$, $\xi_0^{\max}=9.3, 10.5, 12.2$, $N_{\rm elem}=664, 800, 976$
that yield an accuracy of the solutions of an order of the fourth
significant figures.

Figure \ref{nt11}   demonstrates  non-monotonic behavior of the
total transmission probability versus the energy, and the observed
resonances are manifestations of the quantum transparency effect.
With the barrier height increasing, the peaks become narrower, and
their positions shift to higher energies. The multiplet structure
of the peaks in the symmetric case is similar to that in the
antisymmetric case. For three particles, the major peaks are
double, while  for two and four particles, they are single. For
$A=2$ and $\alpha=10,20$, one can observe the additional
multiplets of small peaks.

Figure \ref{nt10} illustrates the energy dependence of the total
transmission probabilities from the exited states. As the energy
of the initial excited state increases, the transmission peaks
demonstrate a shift towards higher energies, the  set of peak
positions keeping approximately the same as for the transitions
from the ground state  and the peaks just replacing each other,
like it was observed in the model calculations \cite{Volya}. For
example, for $A=3$, the position of the third peak for transitions
from the first two states ($E=10.4167$ and $E=10.4156$)  coincides
with the position of the first peak for the transitions from the
second two states ($E=10.4197$ and  $E=10.4298$).

\begin{figure}[t]
  \epsfig{file=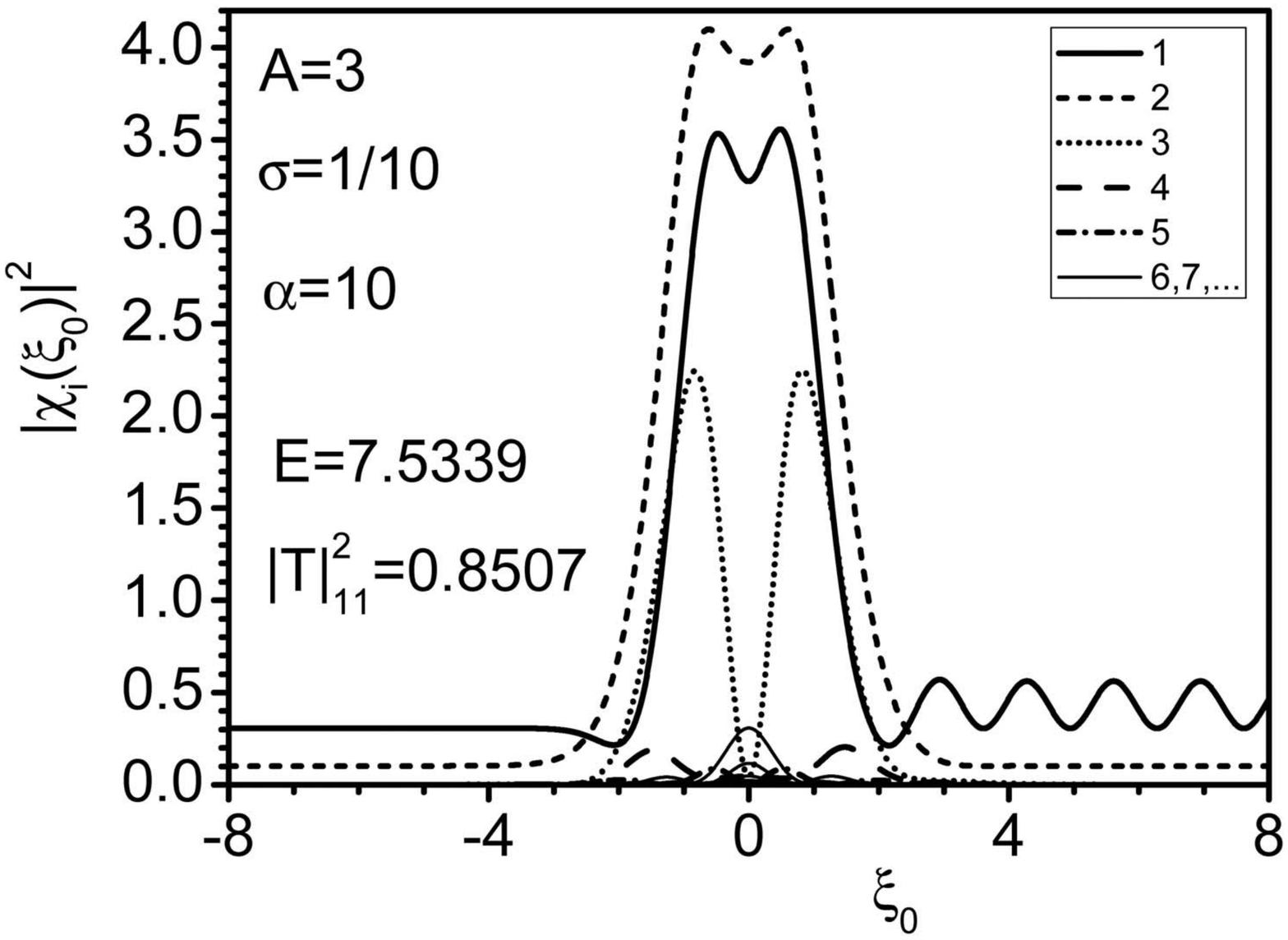,width=0.32\textwidth}
  \epsfig{file=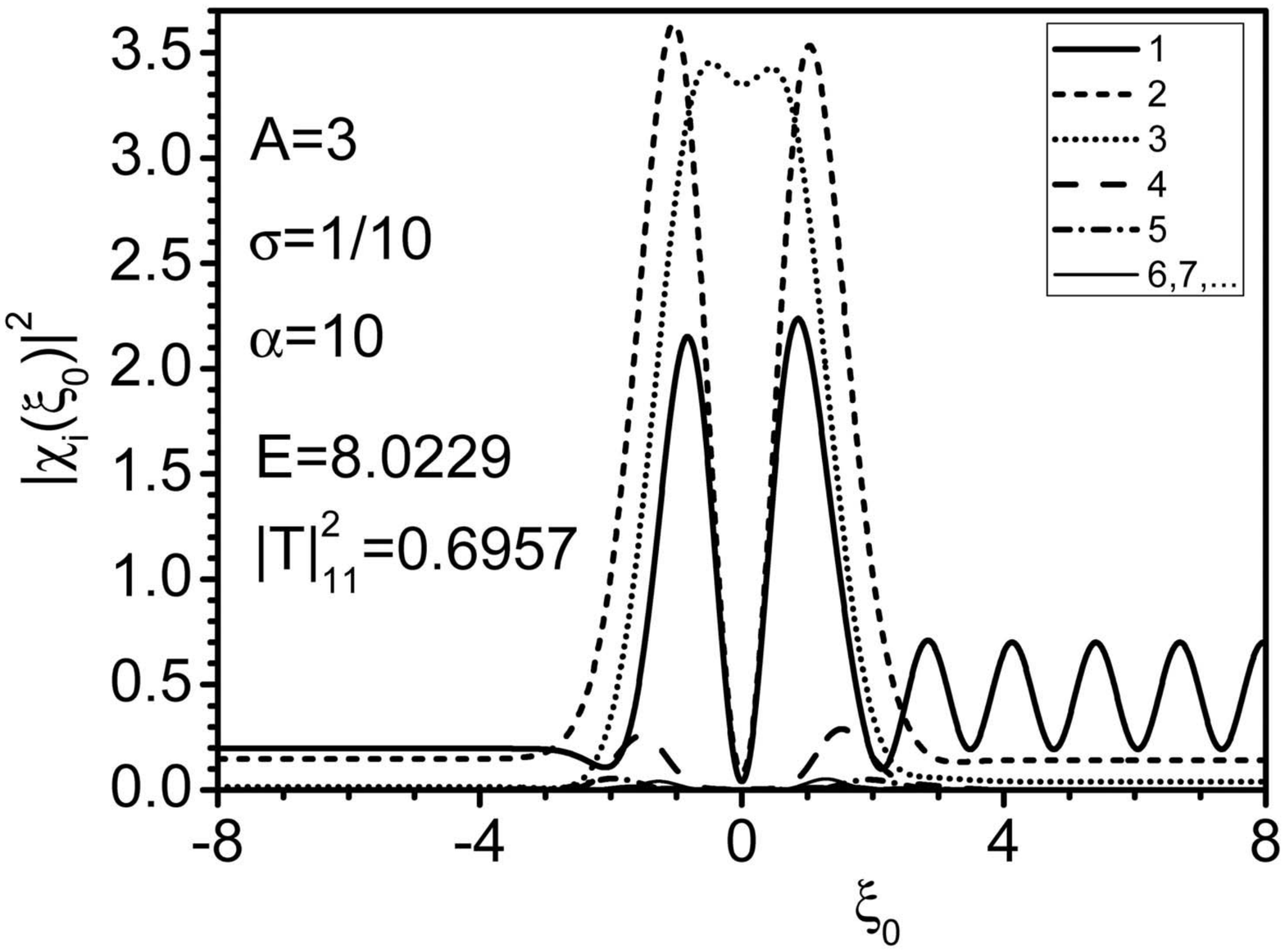,width=0.32\textwidth}
  \epsfig{file=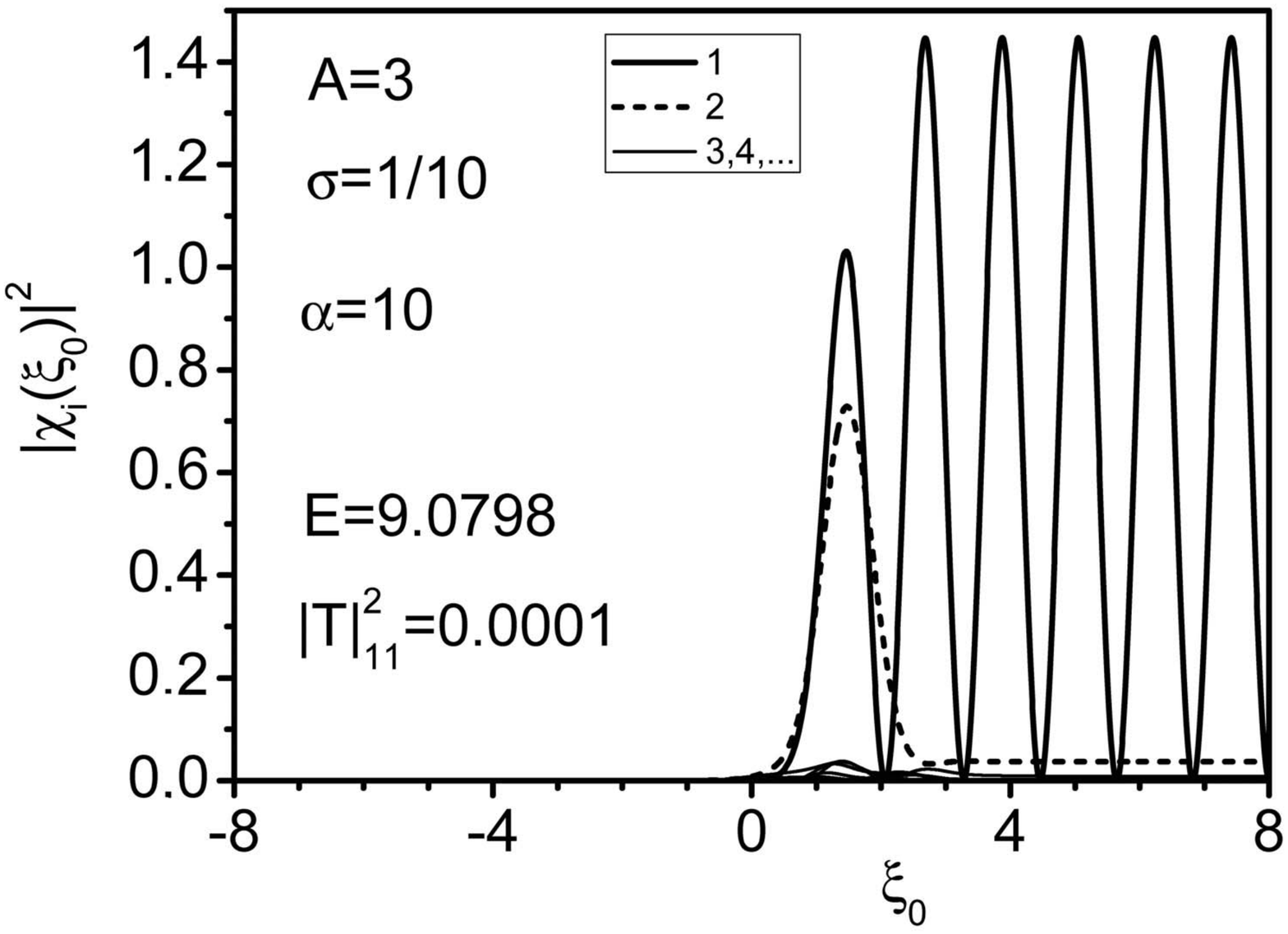,width=0.32\textwidth}\\
  \epsfig{file=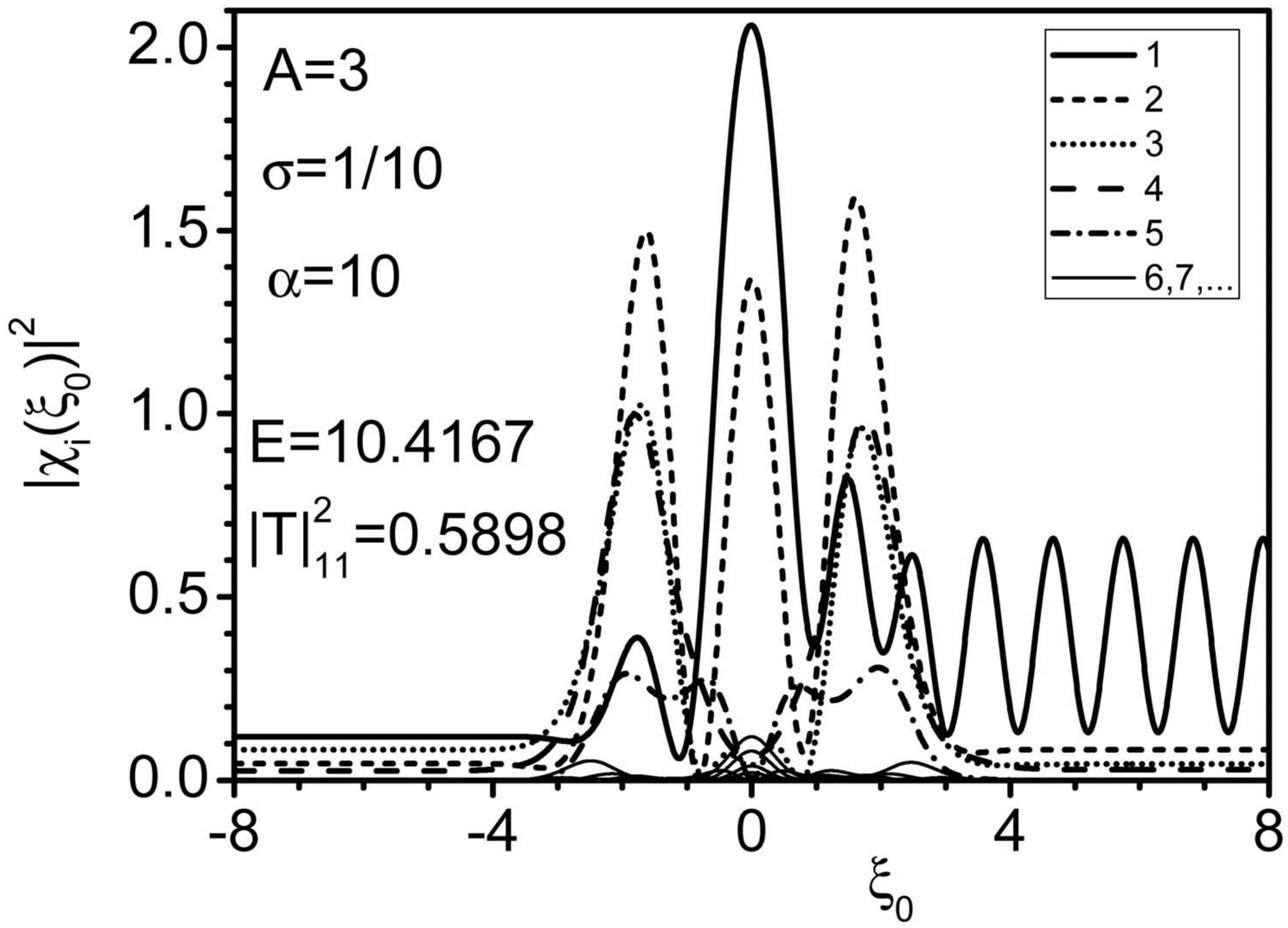,width=0.32\textwidth}
  \epsfig{file=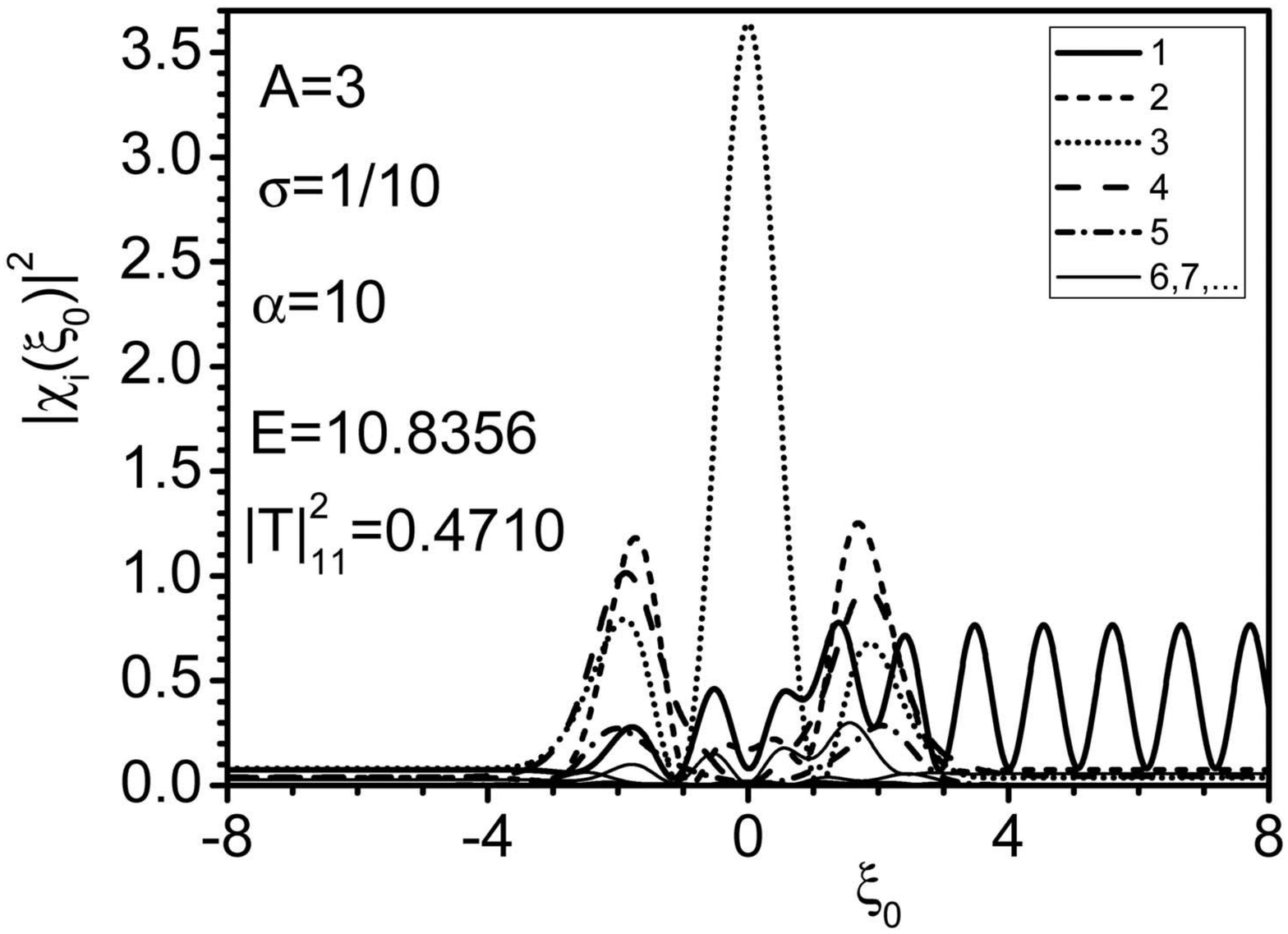,width=0.32\textwidth}
  \epsfig{file=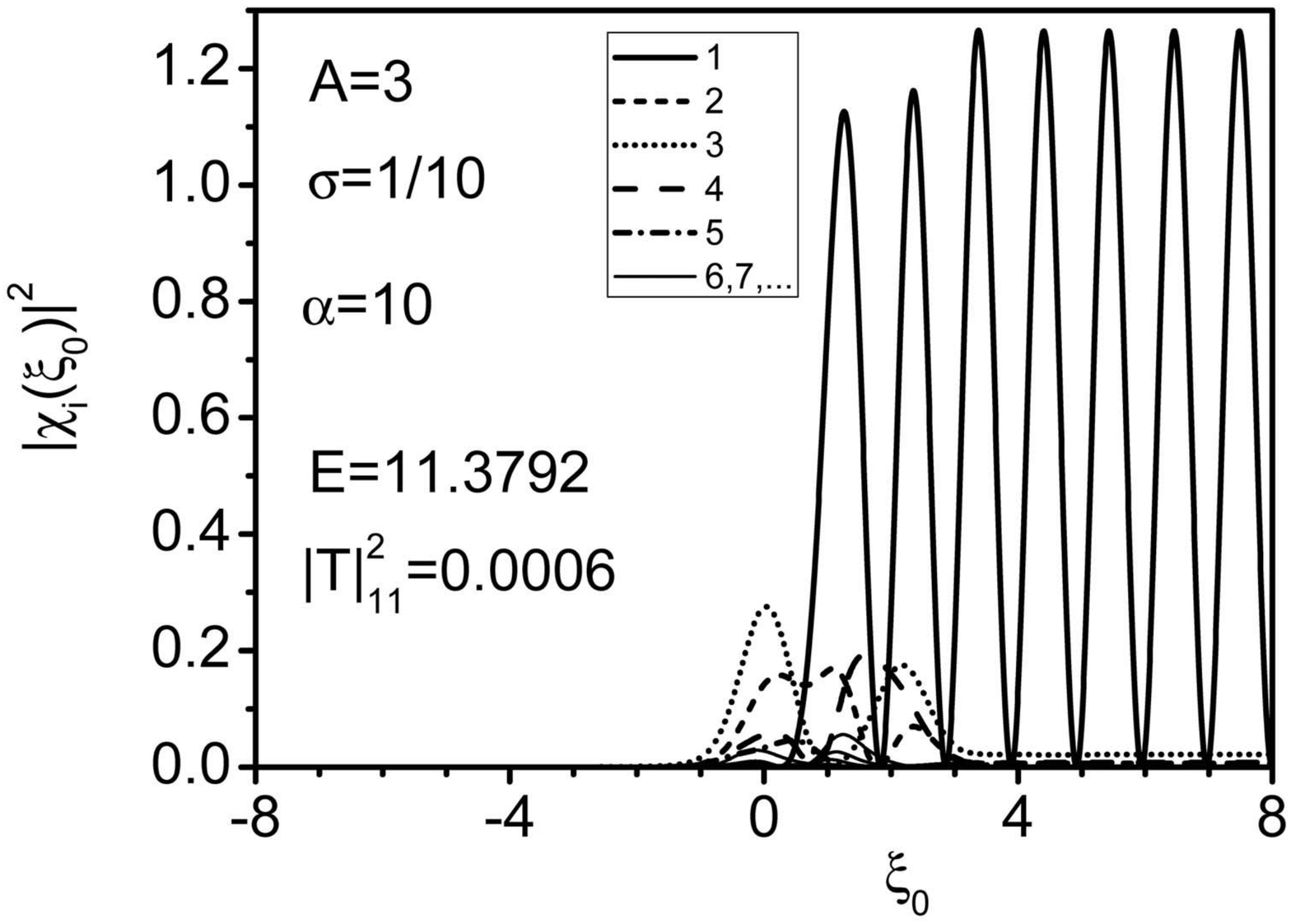,width=0.32\textwidth}\\
  \epsfig{file=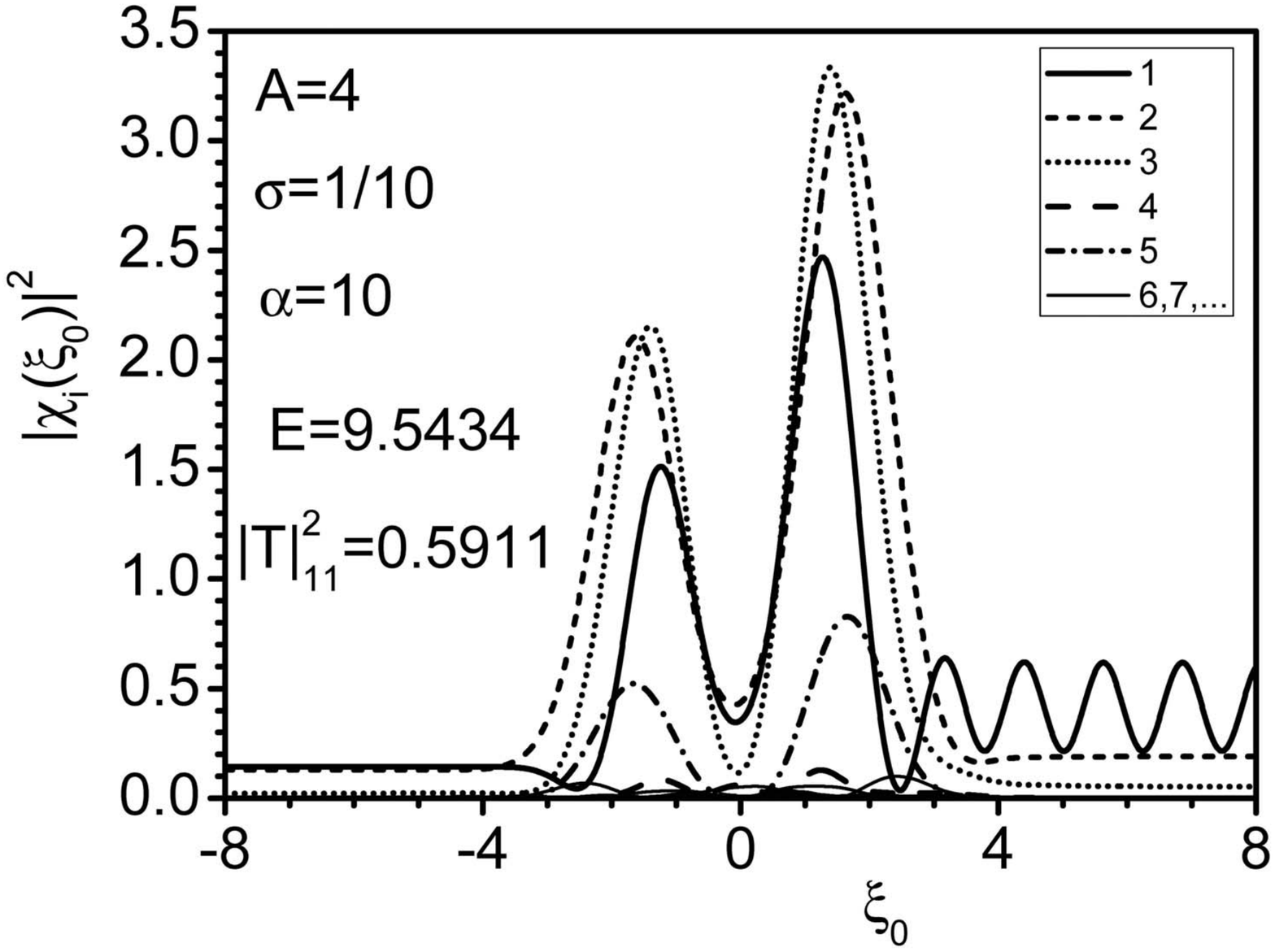,width=0.32\textwidth}
  \epsfig{file=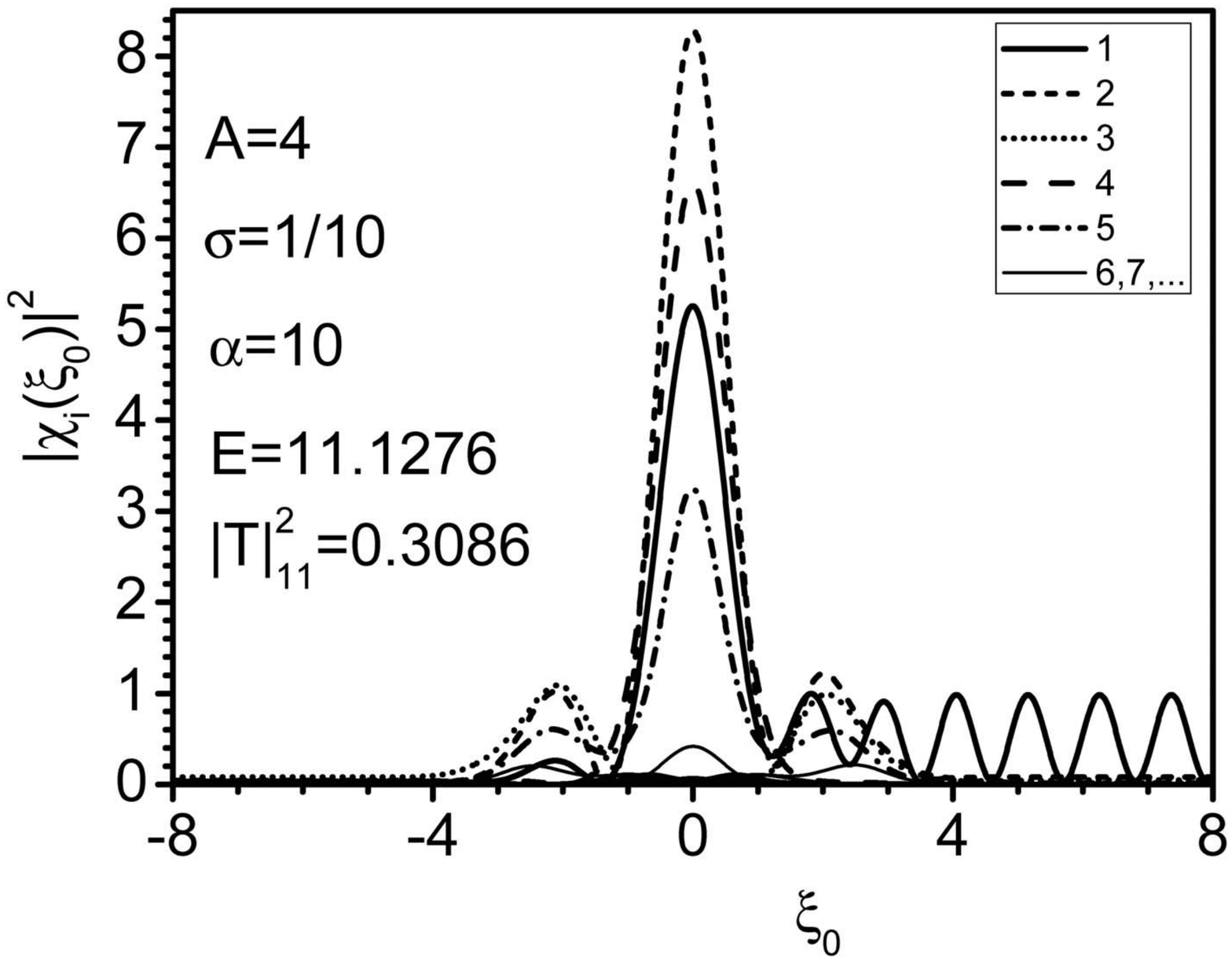,width=0.32\textwidth}
  \epsfig{file=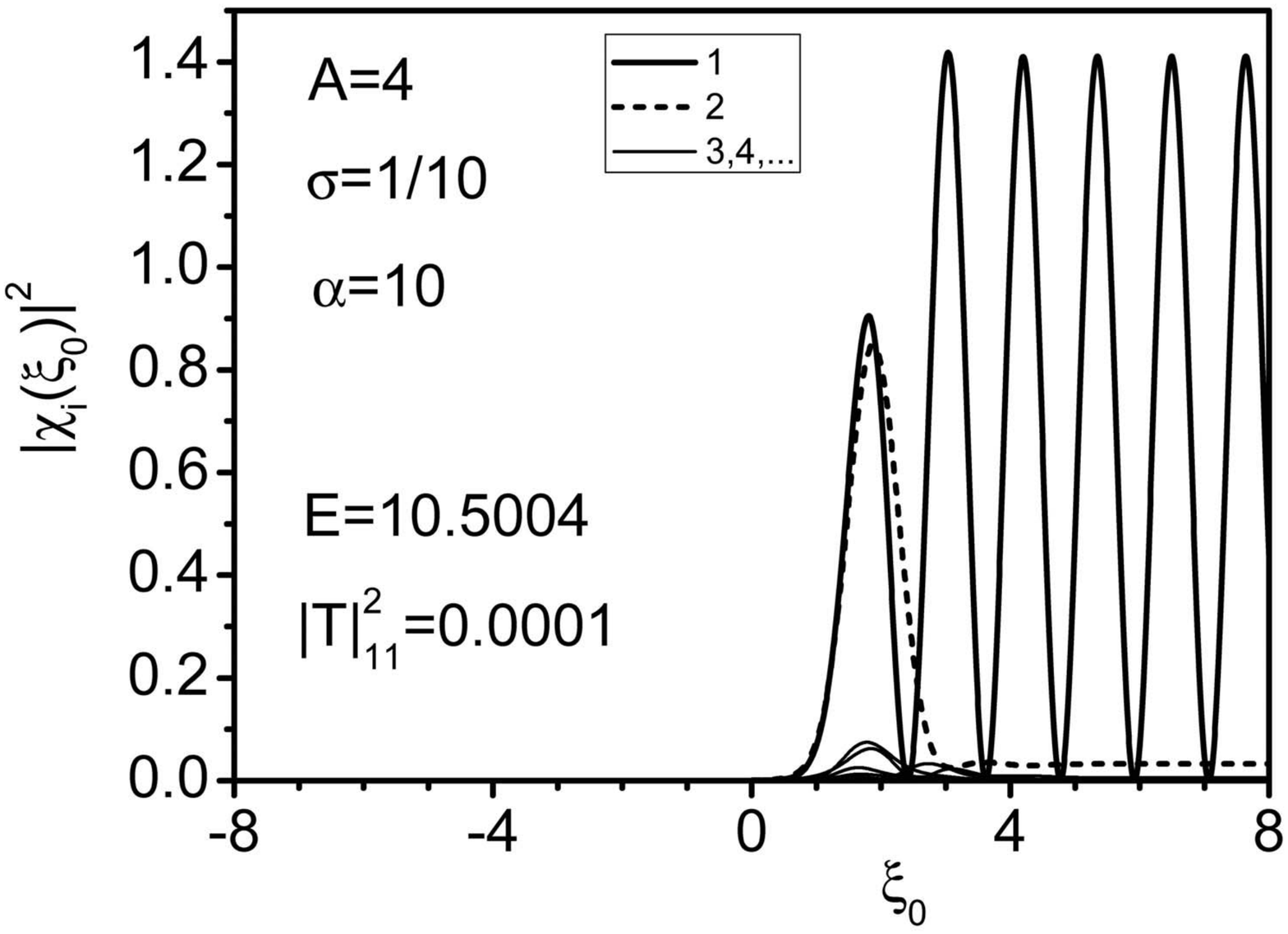,width=0.32\textwidth}
\caption{The probability densities $|\chi_i(\xi_0)|^2$ for the
coefficient functions of the decomposition (\ref{dec}),
representing the incident wave function of the ground  S-state of
the particles at the values of the collision energy $E$
corresponding to individual maxima and minima of the transmission
coefficient in Fig. \ref{nt11}. The parameters of the Gaussian
barrier are $\alpha=10$ and $\sigma=0.1$
 } \label{max}
\end{figure}
{\bf Calculation of energy position of the barrier quasistationary
states\\} In the considered case,  the  potential barrier $V(x_i)$
is narrow, and $V^{pair}(x_{ij})=V^{hosc}(x_{ij})$, so that we
solve Eq. (\ref{mo1})  in the Cartesian coordinates $x_1,...,x_A$
in one of the $2^A-2$ subdomains, defined as $p_ix_i>0$,
$p_i=\pm1$, under the Dirichlet conditions (DC):
$\Psi(x_1,...,x_A)|_{\cup_{i=1}^A \{x_i=0 \}}=0$ at the internal
boundaries $\cup_{i=1}^A \{x_i=0 \}$. Here the value $p_i=\pm 1$
indicates the location of the $i^{th}$ particle at the right or
left side of the barrier, respectively. Thus, in the DC procedure
we seek for the solution in the form of a Galerkin expansion over
the orthogonal truncated oscillator basis, $ \Psi^{D}_{i}( {\bf
x})\!= \sum_{j=1}^{j_{max}}\bar\Phi_{j}({\bf x})\Psi^{D}_{ji} $
composed of $A$-dimensional harmonic oscillator functions
$\bar\Phi_{j}({\bf x})$, odd  in each of the Cartesian coordinates
$x_1,...,x_A$ in accordance with the above DCs, with unknown
coefficients $\Psi^{D}_{ji}$. As a result, we arrive at the
algebraic eigenvalue problem {\boldmath $D\Psi^{\rm D}=\Psi^{\rm
D} {E}^{\rm D}$} with a  dense real-symmetric $j_{max}\times
j_{max}$ matrix. So, in the DC procedure we seek for an
approximate solution in one of the potential wells, i.e., we
neglect the tunnelling through the barriers between wells.
Therefore, we cannot observe the  splitting inherent in exact
eigenvalues corresponding to S and A eigenstates, differing in
permutation symmetry. However, we can explain the mechanism of
their appearance and give their classification, which is
important, too.
This \textit{algorithm DC} was implemented in CAS MAPLE and  FORTRAN environment.\\
{\it Remark.} The DC procedure is similar to solving  Eq. (\ref{mo5}) in the symmetrized coordinates $\xi_0,\mbox{\boldmath$\xi$}$
related to the Cartesian ones by Eq. (\ref{mo2}), implemented the following two steps:\\
({\it i}) we approximate the narrow barriers by impenetrable walls $x_k(\xi_0,\mbox{\boldmath$\xi$})=0$;\\
({\it ii}) we superpose these mutually perpendicular walls with the coordinate hyperplanes using rotations. \\
Actually, the two approaches yield the same boundary-value problem
formulated in different coordinates (\ref{mo1}), (\ref{mo5}).

\textbf{The algorithm \textit{DC}:
}\\[-3mm]
\underline{\hspace{.995\textwidth}}\\
\textbf{Input}:\\
$A$ is the number of identical particles;\\
$x_k$, $k=1,...,A$ are the Cartesian coordinates of the identical particles;\\
$p_k=\pm1$ indicates the location of
the $k^{th}$ particle ;\\ 
$j_{max}$ is the number of the eigenfunctions of A-dimensional harmonic oscillator; \\ [-4mm]
\noindent \underline{\hspace{.995\textwidth}}\\
\textbf{Output}:\\
$D=\{D_{j'j}\}$ is the  $j_{max}\times j_{max}$
matrix ;\\ $E_i^D$ and $\Psi^{D}_{ji}$  are the real-value eigenenergies and eigenvectors;
 \\ [-4mm]
\noindent \underline{\hspace{.995\textwidth}}\\
\textbf{Local}:\\
$\Phi_j=\sqrt{2^A}\prod_{k=1}^A\bar\Phi_{i_k}(x_k)$; \\
$I(i_k',i_k)=\int_0^\infty \bar\Phi_{i_k'}(x)\bar\Phi_{i_k}(x)dx=\frac{2^{(i_k'+i_k)/2}{}_{2}F_{1}(i_k',i_k;(2-i_k'-i_k)/2;1/2)}{\Gamma((2-i_k'-i_k)/2)
\sqrt{i_k'!i_k!}};$ \\
$\Gamma(*)$ is the gamma-function, $_{2}F_{1}(*,*;*;*)$ is the hypergeometric function;\\[-4mm]
\underline{\hspace{.995\textwidth}}\\
1: $Eq:=(-\Delta+\sum (p_kx_k-p_{k'}x_{k'})/2A)$;\\
2: $Eq:=\sqrt{A/(A-1)}(Eq,\Delta \to \Delta/(A/(A-1)), x_k\to x_k\sqrt[4]{A/(A-1)}$;\\
3: $Eq:=Eq, p_k^2\to 1, \Delta =\sum_k(x_k^2- (2n_k+1))$;\\
4: $Eq:=Eq\prod\bar\Phi_{i_k}(x_k)$;\\
5: $Eq:=x_k=(\sqrt{i_k+1}\bar\Phi_{i_k+1}(x_k)+\sqrt{i_k}\bar\Phi_{i_k-1}(x_k))/(\sqrt{2} \bar\Phi_{i_k}(x_k))$;\\
6: \textbf{for}  $j,j'=1,...,j_{max}$ \textbf{do}\\
\phantom{666}  $D_{j'j}:= \Phi_{i_k}(x_k)\to I(i_k',i_k)$;\\
\phantom{6:} \textbf{ end for} \\
7: $D\Psi^{D}_{ji}=\Psi^{D}_{ji}E_i^D$ $\to$ $E_i^D$ and $\Psi^{D}_{ji}$;
\\ [-3mm]
\underline{\hspace{.995\textwidth}}\\[-3mm]

In Table \ref{res2}, we present the resonance values of the energy
$E_S$ ($E_A$) calculated by solving the boundary-value problem
(\ref{mo3e}) and (\ref{new12}), using the KANTBP 3.0 program, for
S (A) states at $A=2,3,4$ $\sigma=1/10$, $\alpha=20$ that
correspond to the maxima of transmission coefficients $|T|^2_{ii}$
in Fig. \ref{nt11} up to values of energy $E<18$ and corresponding
resonance values of the energy $E_D$ calculated by means of the
algorithm DC. One can see that the accepted approximation of the
narrow barrier with impermeable walls using in the algorithm DC
provides the appropriate approximations $E_{i}^{D}$ of the above
high accuracy results $E_S$ ($E_A$) with the error smaller than
2\%. Below we give a comparison and qualitative analysis of the
obtained results.

For two particles, $A=2$ (see Fig. \ref{ga}), there are two
symmetric potential wells. In each of them both symmetric and
asymmetric wave functions are constructed.  Since the potential
barrier separating the wells is sufficiently high, the appropriate
energies are closely spaced, so that each level describes the
states of both S and A type. The lower energy levels form a
sequence ``singlet-doublet-triplet, etc.'', which is seen in Fig.
\ref{nt11}. The resonance transmission energies for a pair of
particles in S states are lower than that for a pair of those in A
states. This is due to the fact that in the vicinity of the
collision point, the wave function is zero.
\begin{figure}[t]
\epsfig{file=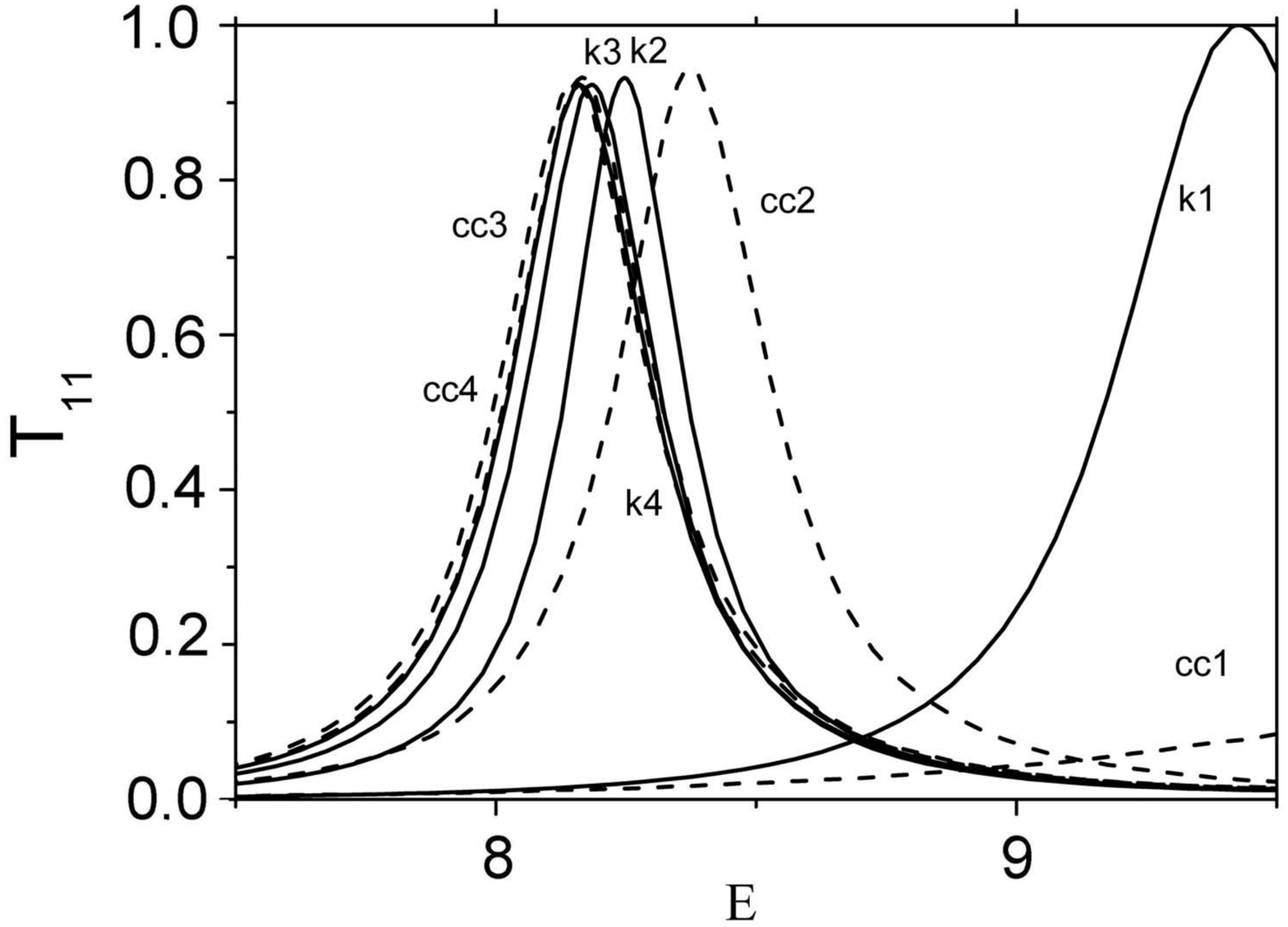,width=0.45\textwidth,height=0.32\textwidth,angle=0}
\epsfig{file=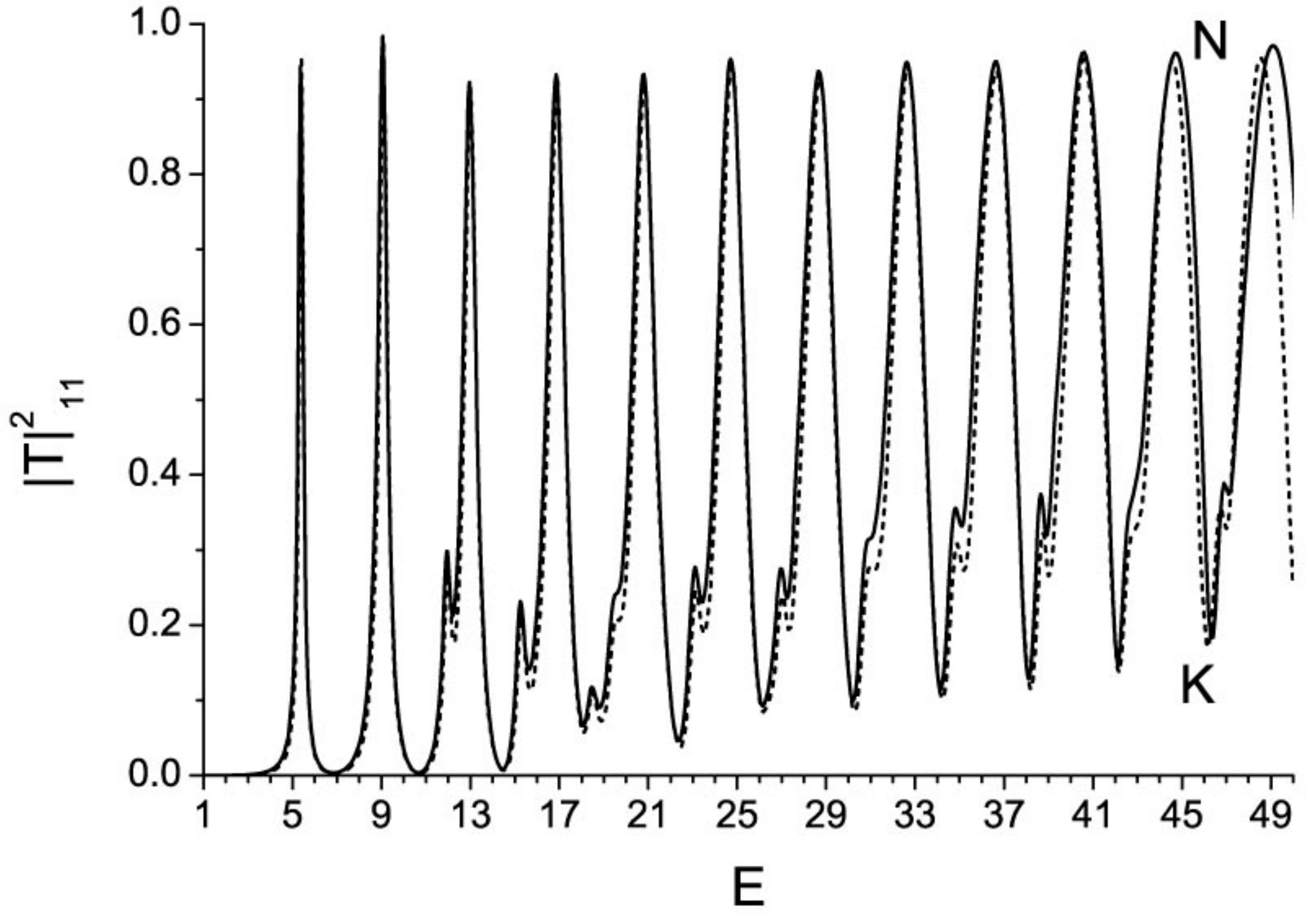,width=0.45\textwidth,height=0.32\textwidth,angle=0}
\caption{a. The  comparison of convergence rate of Galerkin (cc*)
and Kantorovich (k*) close-coupling expansions in calculations of
transmission coefficient $|T|_{11}^2$ for the S-states,  $A=2$ at
$\alpha=10$, $\sigma=0.1$, like epure of the first peak from Fig.
\ref{nt11}. b. The  comparison of  Galerkin and Kantorovich
methods (G=K) with Finite-Difference Numerov method (N)
  }
\label{comp}
\end{figure}
When $A=3$
 there are six similar wells, three of them at each side of the plane $\xi_0=0$.
 The symmetry with respect to the plane
 $\xi_0=0$ explains the presence of doublets. The presence of states
with definite symmetry is associated with the fact that the axis
$\xi_0$ is a third-order symmetry axis. However, in contrast to the
case  $A=2$, one can obtain either S or A combinations of
states. For example, the first four solutions of the problem,
in one of the wells (e.g., the one restricted
with the pair-collision planes   ``13'' and ``23'') possess the
dominant components
 $2\sqrt{2}\bar\Phi_{1}(x_1)\bar\Phi_{1}(x_2)\bar\Phi_{1}(x_3)$,
 $2(\bar\Phi_{1}(x_1)\bar\Phi_{3}(x_2)+\bar\Phi_{3}(x_1)\bar\Phi_{1}(x_2))\bar\Phi_{1}(x_3)$,
{$2(\bar\Phi_{1}(x_1)\bar\Phi_{3}(x_2)-\bar\Phi_{3}(x_1)\bar\Phi_{1}(x_2))\bar\Phi_{1}(x_3)$,}
 $2\sqrt{2}\bar\Phi_{1}(x_1)\bar\Phi_{1}(x_2)\bar\Phi_{3}(x_3)$.
 Note that the first, second, and fourth of these functions are
symmetric with respect to the permutation
 $x_1\leftrightarrow x_2$, while the third one is antisymmetric.
Hence, in all six wells using the first four solutions one can obtain
six S and two A states.

When $A=4$ there are 14 wells. Six wells at the center correspond
to the case when two particles are located at one side of the
barrier and the rest two at the other side. The corresponding
eigenenergy is denoted $E^{D22}_i$. The rest eight wells
correspond to the case when one  particle is located at one side
of the barrier and the rest three at the other side. The
corresponding eigenenergy is denoted $E^{D31}_i$. For these
states, doublets must be observed, similar to the case of three
particles. However, the separation between the energy levels is
much smaller, because the 4-well groups are strongly separated by
two barriers, instead of only one barrier in the case $A=3$.

The necessary condition for the quasi-stationary state being
symmetric (antisymmetric) is that the wave functions must
be symmetric (antisymmetric) with respect to those coordinates $x_i$ and $x_j$,  for which $p_i=p_j$.

The effect of quantum transparency is caused by the existence of
barrier quasistationary states imbedded in the continuum. Fig.
\ref{max} shows that in the case of resonance transmission, the
wave functions depending on the center-of-mass variable $\xi_0$
are localized in the vicinity of the potential barrier center
($\xi_0=0$).

For the energy values corresponding to some of the transmission
coefficient peaks in Fig. \ref{nt11} at $\alpha=10$ within the
effective range of barrier potential action, the wave functions
demonstrate considerable  increase (from two to ten times) of the
probability density in comparison with the incident unit flux.
This is a fingerprint of quasistationary states, which is not a
quantitative definition, but a clear evidence in favor of their
presence in the system\cite{nussenzweig}. In the case of total
reflection, the wave functions are localized at the barrier side,
on which the wave is incident, and decrease to zero within the
effective range of the barrier action.

Note that the explicit explanation of the quantum transparency
effect is achieved in the framework of Kantorovich close-coupling
equations because of the multi-barrier potential structure of the
effective potential, appearing explicitly even in the diagonal or
adiabatic approximation, in particular, in the S case for $A=2$
\cite{P00,casc11}. Nevertheless, in Galerkin close-coupling
equations, the multi-barrier potential structure of the effective
potential is observed explicitly in the A case  (see Fig.
\ref{vij}).

\begin{figure}[t]
  \epsfig{file=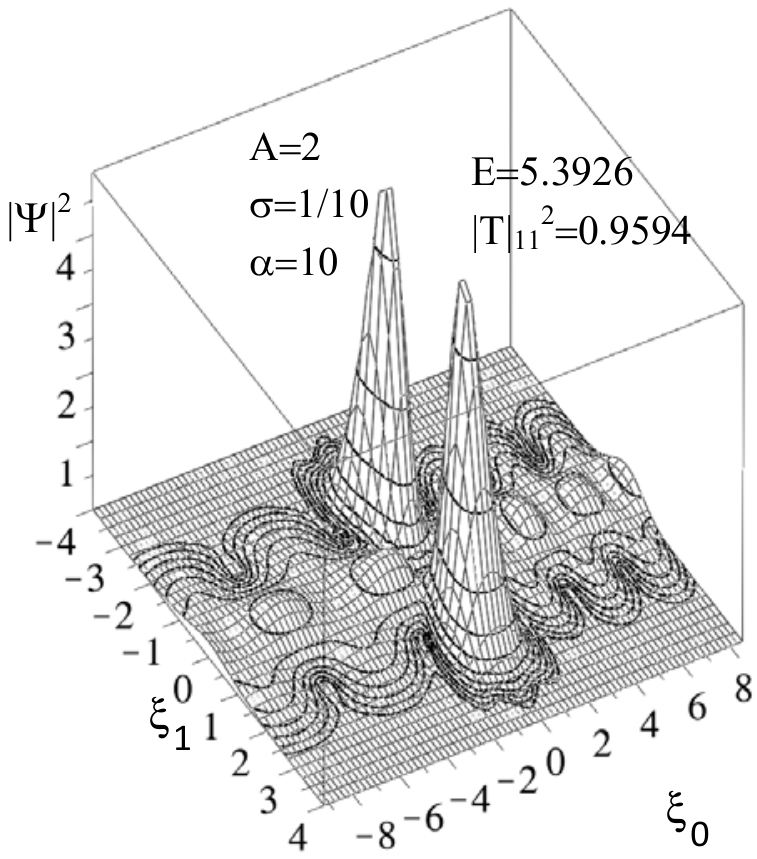,width=0.32\textwidth}
  \epsfig{file=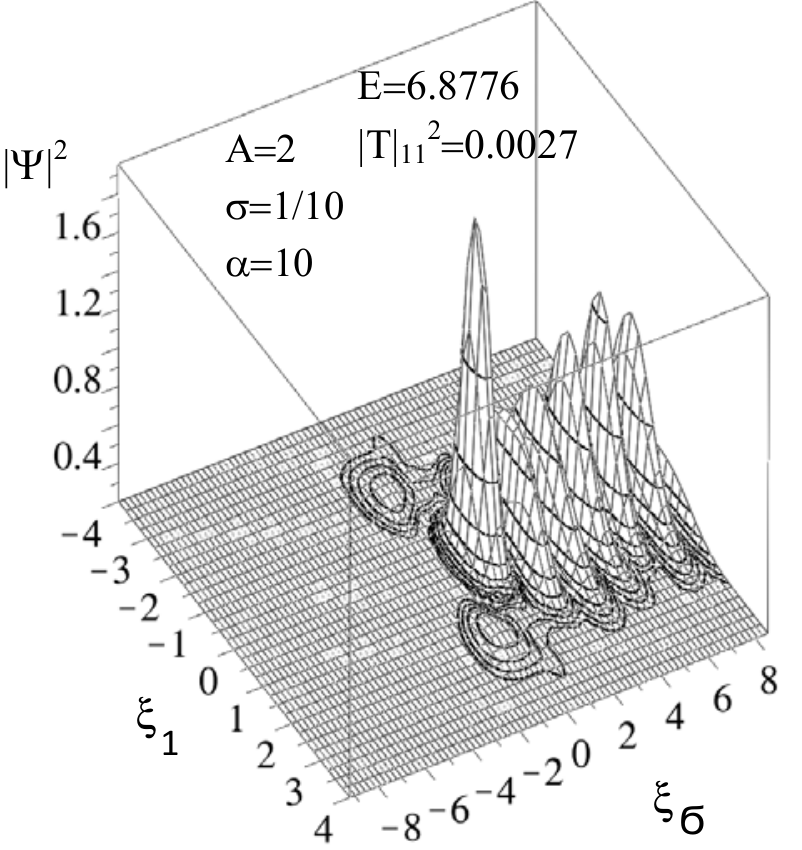,width=0.32\textwidth}
  \epsfig{file=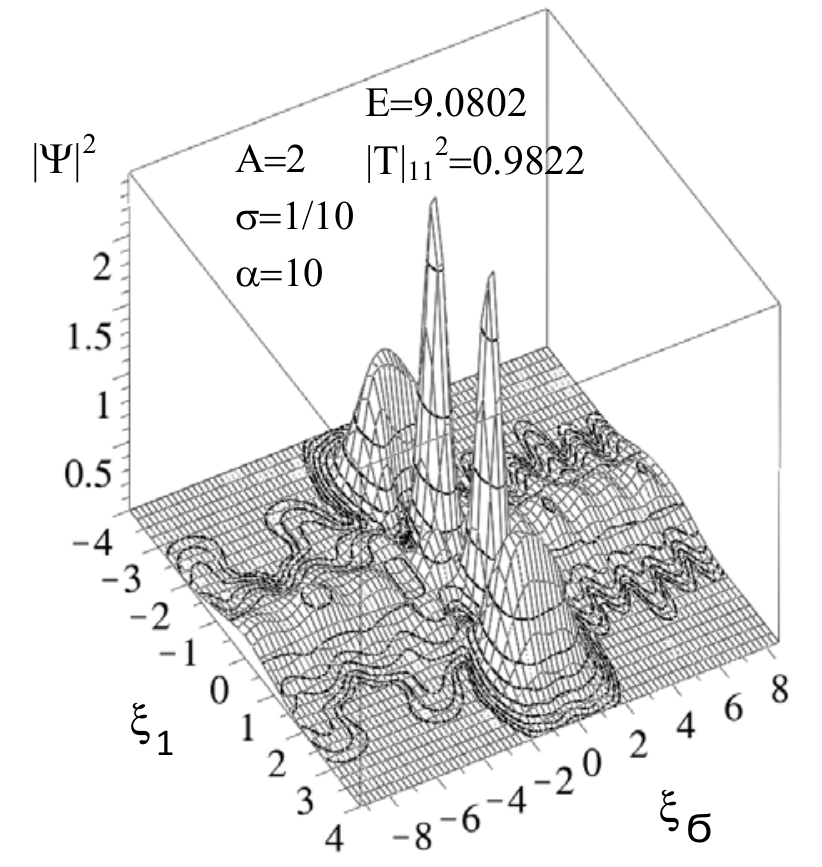,width=0.32\textwidth}
\\
  \epsfig{file=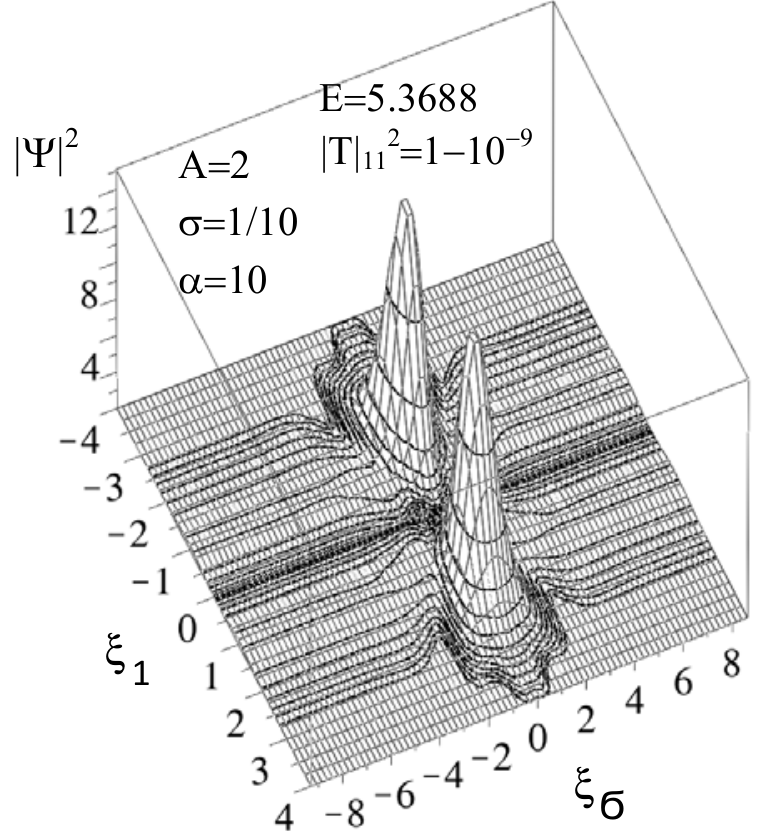,width=0.32\textwidth}
  \epsfig{file=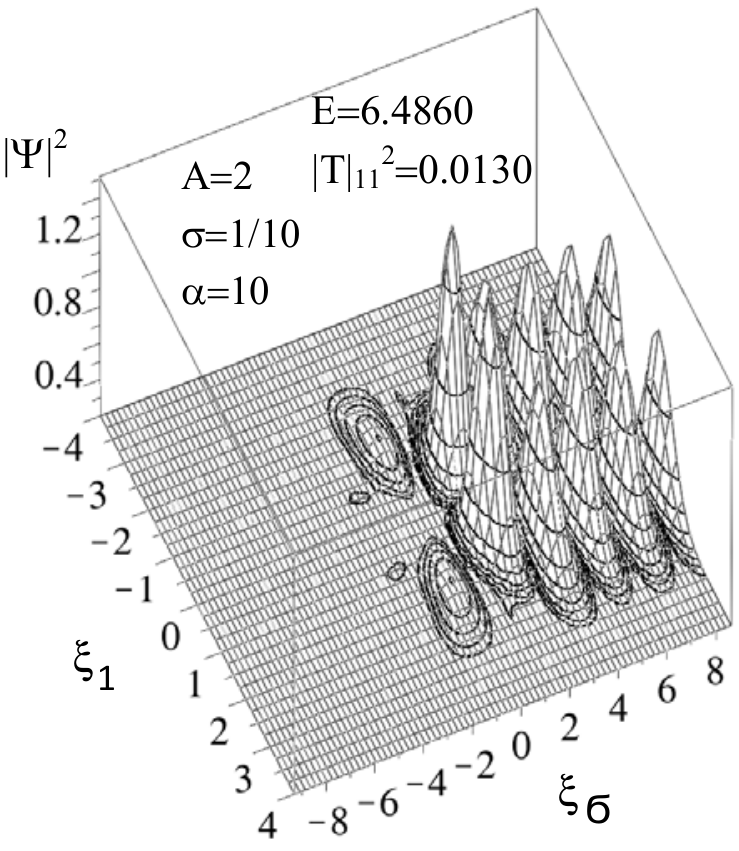,width=0.32\textwidth}
  \epsfig{file=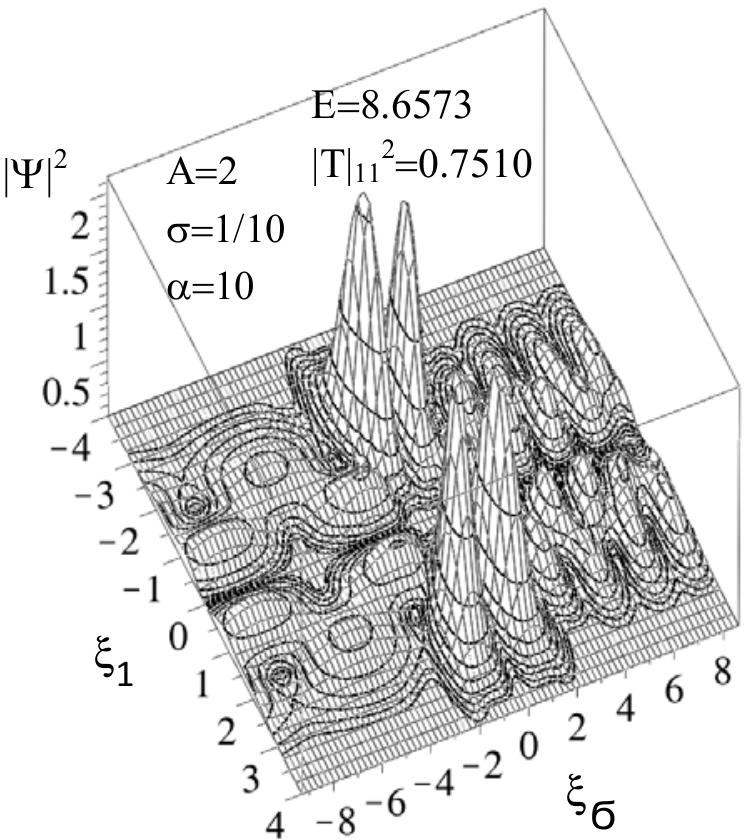,width=0.32\textwidth}
\caption{The profiles of probability densities $|\Psi(\xi_0,\xi_1)|^2$ for the S- (upper panel)
 and A- (lower panel) states of $A=2$ particles, revealing  resonance transmission
 and total reflection at resonance energies, shown in Figs. \ref{nt11}
 } \label{2max}
\end{figure}
As an example,  Fig. \ref{comp}a, which is an epure of Fig.
\ref{nt11}, shows the comparison of convergence rates of Galerkin
(\ref{dec}) and Kantorovich  close-coupling expansions in
calculations of transmission coefficient $|T|_{11}^2$ for S wave functions,
$A=2$ at $\alpha=10$, $\sigma=0.1$. 
One can see that the diagonal approximation  of the Kantorovich
method provides better approximations of the positions of the
transmission coefficient $|T|_{11}^2$ resonance peaks. With the
increasing number of basis functions, i.e., the number $j_{\max}$
of close-coupling equations with respect to the center-of-mass
coordinates in Galerkin (\ref{mo3e}) and Kantorovich  form,
respectively, the convergence rates are similar and confirm the
results obtained by solving the problem by means of the
Finite-Difference Numerov method in 2D domain \cite{P00}, see Fig.
\ref{comp} b. This is true for the considered short-range
potentials (\ref{mo3}), while for long-range potentials of the
Coulomb type, the Kantorovich method can be more efficient
\cite{casc11}.

Figure \ref{2max}  shows the profiles of
$|\mbox{\boldmath$\Psi$}|^2\equiv|\mbox{\boldmath$\Psi$}_{Em\to}^{(-)}|^2$
for the S and A total wave functions of the continuous spectrum in
the $(\xi_0,\xi_1)$ plane with $A=2$, $\alpha=10$, $\sigma=1/10$
at the resonance energies of the first and the second maximum and
the first minimum of the transmission coefficient demonstrating
{\it resonance transmission}  and {\it total reflection},
respectively. It is seen that in the case of resonance
transmission, the redistribution of energy from the center-mass
degree of freedom to the internal (transverse) ones takes place,
i.e., the transverse oscillator undergoes a transition from the
ground state to the excited state, while in the total reflection,
the redistribution of energy is extremely small, and the
transverse oscillator returns to infinity in the same state.

\section{Conclusion}
We considered a model cluster of $A$ identical particles bound by
the oscillator-type potential that undergo  quantum tunnelling
through the short-range repulsive barrier potentials. The model
was formulated in the new  representation, which we referred as
the Symmetrized Coordinate Representation (SCR, see forthcoming
paper \cite{cascnedve1}), that implies construction of symmetric
(asymmetric) combinations of oscillator wave functions in new
coordinates. The approach was implemented as a  complex of the
symbolic-numeric algorithms and programs.

For clarity, a  system of several identical particles was
considered in one-dimensional Euclidian space ($d=1$). We
calculated only the spatial part of the wave function, symmetric
or antisymmetric under permutation of $A$ identical particles. If
necessary, the spin part of the wave function can be introduced
using the conventional procedure for more rigorous calculation.

We analyzed
 the effect of quantum transparency, i.e., the resonance tunnelling of several bound particles
 through repulsive potential barriers.
 We demonstrated  that this effect is due to the existence of sub-barrier quasistationary states imbedded in the continuum.
 For the considered type of symmetric Gaussian barrier potential, the energies of the
 S and A quasistationary states
 are slightly different
 because of the similarity of the multiplet structure of oscillator energy levels at a fixed number of particles.
This fact explains  a similar behavior of transmission
coefficients for S and A states shifted by threshold energies. The
multiplet structure of these states is varied with increasing the
number of particles, e.g., for three particles, the major peaks
are double, while for two and four particles, they are single. Our
calculations have also shown that with increasing the energy of
the initial excited state of few-body clusters, the transmission
peaks demonstrate a shift towards higher energies, the  set of
peak positions keeping approximately the same as for the
transitions from the ground state  and the peaks just skipping
from one position to another.

The proposed approach can be adapted and applied to
tetrahed\-ral-symmetric nuclei,  quantum diffusion of molecules and
micro-clusters through surfaces,
and
fragmentation mechanism in producing very
neutron-rich light nuclei.
In connection with the intense search for superheavy nuclei, a particularly significant application of the proposed approach is the mathematically correct analysis of mechanisms of  sub-barrier fusion of heavy nuclei and the study of fusion rate enhancement by means of resonance tunnelling.

The authors thank Professors V.P. Gerdt, A. G\'o\'zd\'z, and F.M. Penkov for collaboration. The work was supported  by grants 13-602-02 JINR, 11-01-00523 and 13-01-00668  RFBR, 0602/GF MES RK
and the Bogoliubov-Infeld program.

\end{document}